\documentclass[prc,twocolumn,showpacs,superscriptaddress]{revtex4}

\usepackage{graphicx}
\usepackage{epsfig}
\usepackage{dcolumn}% Align table columns on decimal point
\usepackage{bm}% bold math
\usepackage{amssymb}

\begin{document}

\preprint{Preprint}

\title{Vector and Tensor Analyzing Powers of the \\
\mbox{$\mathrm{^1H(\vec{d},\gamma)^3He}$}--capture reaction}

\author{T.~Klechneva}\affiliation{Dept. f\"ur Physik und Astronomie, Universit\"at Basel, Switzerland} \affiliation{St.Petersburg Nuclear Physics Institute, Gatchina, Russia}
\author{C.~Carasco}\affiliation{Dept. f\"ur Physik und Astronomie, Universit\"at Basel, Switzerland}
\author{I.~Goussev}\affiliation{St.Petersburg Nuclear Physics Institute, Gatchina, Russia}
\author{M.~Hauger}\affiliation{Dept. f\"ur Physik und Astronomie, Universit\"at Basel, Switzerland}
\author{J.~Jourdan}\email{Juerg.Jourdan@unibas.ch}\affiliation{Dept. f\"ur Physik und Astronomie, Universit\"at Basel, Switzerland} 
\author{B.~Krusche}\affiliation{Dept. f\"ur Physik und Astronomie, Universit\"at Basel, Switzerland}
\author{H.~M\"uhry}\affiliation{Dept. f\"ur Physik und Astronomie, Universit\"at Basel, Switzerland} 
\author{Ch.~Normand}\affiliation{Dept. f\"ur Physik und Astronomie, Universit\"at Basel, Switzerland}
\author{D.~Rohe}\affiliation{Dept. f\"ur Physik und Astronomie, Universit\"at Basel, Switzerland}
\author{D.~Seliverstov}\affiliation{St.Petersburg Nuclear Physics Institute, Gatchina, Russia}
\author{I.~Sick}\affiliation{Dept. f\"ur Physik und Astronomie, Universit\"at Basel, Switzerland}
\author{G.~Testa}\affiliation{Dept. f\"ur Physik und Astronomie, Universit\"at Basel, Switzerland}
\author{G.~Warren}\affiliation{Dept. f\"ur Physik und Astronomie, Universit\"at Basel, Switzerland}
\author{H.~W{\"o}hrle}\affiliation{Dept. f\"ur Physik und Astronomie, Universit\"at Basel, Switzerland}
\author{M.~Zeier}\affiliation{Dept. f\"ur Physik und Astronomie, Universit\"at Basel, Switzerland}

\date{\today}% It is always \today, today,

\begin{abstract}
Precise measurements of the deuteron vector analyzing power A$_y^d$ and the tensor analyzing power A$_{yy}$
of the $\mathrm{^1H(\vec{d},\gamma)^3He}$--capture reaction have been performed at deuteron
energies of 29~MeV and 45~MeV. The data have been compared to theoretical
state--of--the--art calculations available today. Due to the large sensitivity of polarization 
observables and the precision of the data light could be shed on small effects present in the dynamics 
of the reaction.
\end{abstract}

%\keywords{Radiative Capture, Analyzing Power, Few Body}

\maketitle

% main text
%%%%%%%%%%%%%%%%%%%%%%%%%%%%%%%%%%%%%%%%%%%%%%%%%%%%%%%%%%%%%%%%

\section{Introduction}
\label{intro}
Experimental studies of the three--body system are particularly
interesting as it has become possible to solve the Schr\"odinger equation of the three--body 
system for both the ground-- and the continuum states. Such three--body calculations, based on different techniques, are available from several groups \cite{Gloeckle96,Golak00a,Kievsky94,Kievsky01,Deltuva03b,Yuan02}. Data for d--p elastic scattering, two--, and three--body breakup, and radiative capture reactions (or its inverse reaction, photo-disintegration) now allow for a precise quantitative comparison with theory.  

Combined with these calculations the radiative capture reaction provides 
an especially attractive framework as the electromagnetic interaction is a well understood process.
Capture reactions in few--body systems have been studied for quite some
time and provided, even at very low energy \cite{Bethe50}, valuable information on the dynamics in these systems. In the two--body system the cross section data at low energy indicated
the importance of mesonic degrees of freedom \cite{Riska72} and the
forward angle cross section of the two--body photodisintegration could only be
understood when accounting for relativistic effects \cite{Cambi82}.  

The important role of of meson exchange currents (MEC) can be observed in many 
electromagnetic observables. The electromagnetic form 
factors of the A=3 system as measured in elastic electron scattering \cite{Amroun94} or the electrodisintegration cross section of the two body system \cite{Mathiot84}
can only be understood when MEC's are taken into account. In these observables MEC's 
are intimately related to the nucleonic S--D transitions as they give effects of 
similar size, but opposite sign. Thus, the quantitative study of MEC--effects requires a precise knowledge of the nucleonic S--D transition. The effect of such transitions can be enhanced in measurements of polarization observables as the S--S amplitude, which usually dominates unpolarized cross sections, is strongly suppressed and allows for a study of the small amplitudes. Based on the following arguments one can
expect that measurements of polarization observables in capture reactions provide insight into the
different roles played by nucleonic and mesonic degrees of freedom. 

In the energy range of 10-50 MeV the $\mathrm{^1H(\vec{d},\gamma)^3He}$--capture process is dominated by the electric dipole transition (E1). Although MEC's can give large contributions to the E1-transition, they can be taken into account implicitly when performing calculations with operators derived using Siegert's theorem. Thus, due to the angular dependence of the E1-E1 contribution (sin$^2(\theta_{c.m.})$), polarization observables at medium range reaction angles ($\theta_{c.m.}$) are little affected by the explicit contribution of MEC's. On the other hand MEC's have to be calculated explicitly in the magnetic transitions, particularly M1 in which they may give contributions up to 50\% \cite{Jourdan86}. Contributions of M1 and thus MEC's are particularly large at small and large reaction angles. At these angles the E1-M1 interference term, in which the small M1 amplitude is enhanced via E1 becomes dominant.  Thus, angular distributions of polarization observables with sensitivities to both the electric-- and the magnetic transitions can be expected to offer excellent windows on small components of the interaction and provide a unique experimental test of MEC's.

Measurements of polarization observables for the $\mathrm{^1H(\vec{d},\gamma)^3He}$--capture reaction with adequate precision are rather scarce. Difficulties associated with polarized beam and/or target production, and with the measurement of spin observables often led to experimental uncertainties which do not
permit a significant check of theoretical predictions. Although the techniques are well
under control today, one can not expect a significant increase of the data base as the
required experimental facilities are no longer at hand \cite{Arenhoevel04}. A rather complete
account of the existing data is given in the publication by Anklin
{\em et al.} \cite{Anklin98} which also discusses the results of our previous work in this area. 

\begin{figure}[htb]
\includegraphics[width=0.49\textwidth,clip]{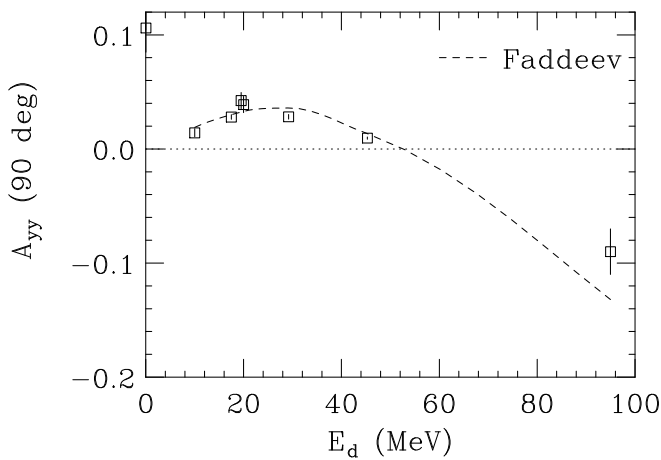}
\caption{\label{dpg_ayy_ed}
Beam energy dependence of A$_{yy}$ at $\theta_{c.m.} = 90$deg. The data points are from \protect{\cite{Schmid96,Sagara95,Vetterli85,Jourdan86,Anklin98,Pitts88}}. The dashed line represents the Faddeev calculation by Golak {\em et al.} \cite{Golak00a}}
\end{figure}

Here we report on new measurements of vector-- and tensor analyzing powers of the $\vec{d}-p$ capture
reaction induced with a polarized deuteron beam. In the present experiment we have extended the measurements by Anklin {\em et al.} in three ways. One extension concerns more extreme angles at a deuteron beam energy of 45~MeV for a larger sensitivity to the magnetic transitions. A second extension concerns the measurement of a complete angular distribution of the polarization observables at a beam energy of 29~MeV. This is particularly important for 
the tensor analyzing power A$_{yy}$ at intermediate angles as the beam energy dependence shows a maximum at 29~MeV while at 45~MeV it is close to zero (see figure \ref{dpg_ayy_ed}). As will be discussed in the last section the tensor analyzing power at these two energies has rather different sensitivities to the underlying physics. As a third extension we also measured the data of the deuteron vector analyzing power A$_y^d$ at the same kinematical points.

\label{exp}
\section{Experiment}
The experiment was performed at the Philips injector cyclotron of the Paul Scherrer Institute (PSI) 
in Villigen (Switzerland). It used a polarized deuteron beam which was prepared in the PSI 
atomic beam ion source \cite{Singy90}. The source was equipped with a 30K cold atomic beam dissociator, 
two sextupole fields to focus (defocus) the atomic electrons from $^2$H with spin up (down), a set of 
two strong (SF1, SF2) and one weak field radio frequency (RF) --transition units to induce the nuclear polarization, 
and an electron--cyclotron resonance (ECR) ionizer. The RF--units selected transitions 
between different Zeemann levels of the $^2$H-atom hyperfine structure in an external magnetic field. 
Depending on the combination of active RF--units the nominal nuclear vector and tensor polarization as 
listed in table \ref{tab:pol} could be prepared. In the present experiment all five modes have been used, 
the source being cycled through them with a rate of 0.3~Hz to minimize systematic errors.

\begin{table}[htb]
\begin{ruledtabular}
\begin{tabular}{cccccc}
Pol. state & SF2 & SF1 & WF & $\hat{p}^i_{z}$ & $\hat{p}^i_{zz}$ \\ [1mm]
\hline 
  a &         off & off & off & $   0$    & $  0$ \\
  b &         on  & off & off & $+1/3$    & $ +1$ \\
  c &         off & on  & off & $+1/3$    & $ -1$ \\
  d &         on  & off &  on & $-1/3$    & $ -1$ \\
  e &         off & on  &  on & $-1/3$    & $ +1$ \\ 
\end{tabular}
\end{ruledtabular}
\caption{\label{tab:pol} Polarized beam source modes and nominal polarization values.}
\end{table}

\begin{figure}[htp]
\includegraphics[width=0.48\textwidth]{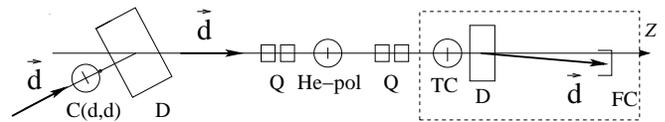}
\caption{\label{fig:setup}Schematic overview of the beam line in the experimental hall NE-C. 
	Here C(d,d) -  carbon scattering chamber,
	Q -  quadrupole doublet, He-pol - $^{4}$He-polarimeter, TC - target chamber, D - dipole magnet 
	to separate  $^{3}$He and unscattered deuterons, FC - Faraday cup.}
\end{figure}

Following the source, the beam was deflected into the injector cyclotron, accelerated to energies 
of 29~MeV and 45~MeV, respectively, and guided to the experimental area. A schematic view of the experimental 
setup in the area is shown in Figure \ref{fig:setup}.

The beam passed first a scattering chamber, where deuterons scattered elastically from a 
thin carbon foil. The distribution of their time of flight relative to the radio-frequency (RF) signal 
from the cyclotron was measured using a fast plastic scintillation detector placed at 30$^{o}$ below the 
beam axis. The cyclotron was tuned for a minimal beam burst width of typically 1.5~ns FWHM. 

In a second target chamber (He-pol), a polarimeter was employed for measurements of the polarization
of the beam at regular intervals. Details of the polarization determination will be explained in section \ref{sec:polarimetry}. Finally the beam was refocused onto a liquid hydrogen target mounted in the third scattering chamber (TC) for the measurement of the 
capture reaction. Downstream of the main target station a C--shaped dipole magnet (D) separated the beam from recoil particles of the capture reaction. A Faraday cup (FC) stopped the beam and measured the current. 
\subsection{Polarimetry} \label{sec:polarimetry}

\begin{figure*}[htp]
\centering
\includegraphics[width=0.75\textwidth]{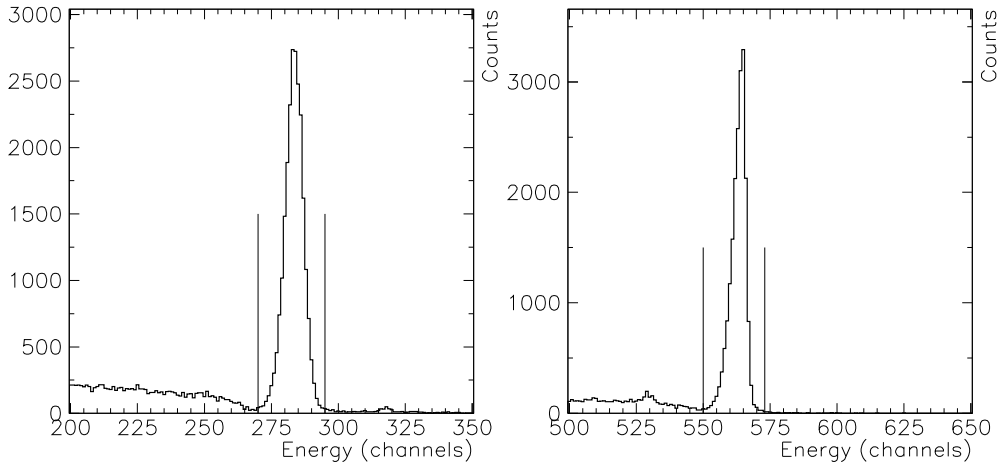}
\caption{\label{fig:pol1}Energy spectra at $\vartheta_{\alpha~lab}$ = 15$^{o}$, $E_{d}=$29~MeV
	(left) and $E_{d}=$45~MeV (right).
	The integration limits are shown as vertical lines.}
\end{figure*}  

Elastic $^4$He($\vec{d},\alpha )$-scattering was used to measure the absolute deuteron beam polarization. 
In the energy range of this experiment at a center of mass angle ($\vartheta_{c.m.}$) of 150 deg the 
analyzing powers are high and precisely known \cite{Anklin98c}. In table \ref{ay_pol} the specific values
used here for the polarization determination of the beam are listed.
\begin{table}[htb]
\begin{ruledtabular}
\begin{tabular}{ccc}
Beam Energy & A$_{y}$ & A$_{yy}$ \\ [1mm]
\hline
29~MeV & $0.846 \pm 0.020$ & $0.910 \pm 0.016$ \\
45~MeV & $0.497 \pm 0.011$ & $0.921 \pm 0.013$ \\
\end{tabular}
\caption{Vector (A$_{y}$) and tensor (A$_{yy}$) analyzing powers for elastic $\vec{d}-\alpha$
scattering at a center of mass angle of 150 deg.\label{ay_pol}}
\end{ruledtabular}
\end{table}

A $^4$He gas cell with 5~$\mu$m thin Havar windows, operating at a pressure of 50~kPa, served as the target
of the polarimeter. At laboratory angles of $\pm 15^{o}$ corresponding to $\vartheta_{c.m.} = \pm 150$deg 
recoil $\alpha$--particles were detected with two symmetrically arranged ``passivated implanted planar 
silicon'' (PIPS) detectors. A double slit system with Ta-collimators at distances of 12~cm and 45~cm from 
the center of the scattering chamber shielded background particles from 
beam--target window reactions. The collimators limited the angular acceptance to $\pm$0.5 deg.
PIPS detectors with thicknesses of 700~$\mu$m (500~$\mu$m) at beam energies of 45~MeV (29~MeV) allowed for 
the discrimination between the stopped recoil $\alpha$--particles and the high energy elastically 
scattered deuterons which deposited less energy as compared to the stopped $\alpha$--particles.
Figure \ref {fig:pol1} displays two sample energy spectra measured at the two deuteron energies. 

The polarization has been determined with the expression for the cross section of a reaction with a
mixed--polarization deuteron beam: 
\begin{equation}
\left( \frac{d \sigma}{d \Omega} \right)^i = \left( \frac{d \sigma}{d \Omega} \right)_a
\left( 1 + \frac{3}{2} \hat{p}^i_z A_y + \frac{3}{2} \hat{p}^i_{zz} A_{yy} \right)
\end{equation}
with $i=b,c,d,e$. Vector (tensor) polarizations of the beam for source state $i = b,c,d,e$ are denoted with 
$\hat{p}^i_z$ ($\hat{p}^i_{zz}$)  (see table \ref{tab:pol}) and $\left(d \sigma/d \Omega \right)_a$ is the 
unpolarized cross section. Coordinate system and symbol definitions follow the Madison convention 
\cite{Barschall71}. Based on the general expression the following equations can then be written for 
each polarization state $i$:
\begin{eqnarray}
\hat{p}^i_{zz} = (\frac{K_{+}^{i}N_{+}^{i}-N_{+}^{0}}{N_{+}^{0}} + \frac{K_{-}^{i}N_{-}^{i}-N_{-}^{0}}
{N_{-}^{0}}) \frac{1}{A_{yy}} \nonumber \\
\hat{p}^i_z = (\frac{K_{+}^{i}N_{+}^{i}-N_{+}^{0}}{N_{+}^{0}} - \frac{K_{-}^{i}N_{-}^{i}-N_{-}^{0}}
{N_{-}^{0}}) \frac{1}{3A_{y}} \label{pol_pol}
\end{eqnarray} \\
with $N_{+}^{i}$ ($N_{-}^{i}$) the number of events detected by the left (right) detector and
$K_{+}^{i}$ ($K_{-}^{i}$) correction factors which account for dead time (DT) and Faraday cup (FC) 
differences for the different states $i$. Cross sections can be replaced by accumulated counts 
because solid angles and efficiencies cancel.

The polarization of the deuteron beam was measured every 8 hours. As 
already observed in the previous experiment by Anklin {\em et al.} \cite{Anklin98} the polarizations 
of the source have been very constant over a run period of 14 days. Typical 
mean polarization values were $\hat{p}^i_z = 0.25$ for the vector and 
$\hat{p}^i_{zz} = 0.65$ for the tensor polarizations for states $i=b-d$ with accuracies of 2--3\%.

\subsection{Setup for the \mbox{$\mathrm{^1H(\vec{d},\gamma)^3He}$} analyzing power measurements}

\begin{figure}[htp]
\includegraphics[width=0.48\textwidth]{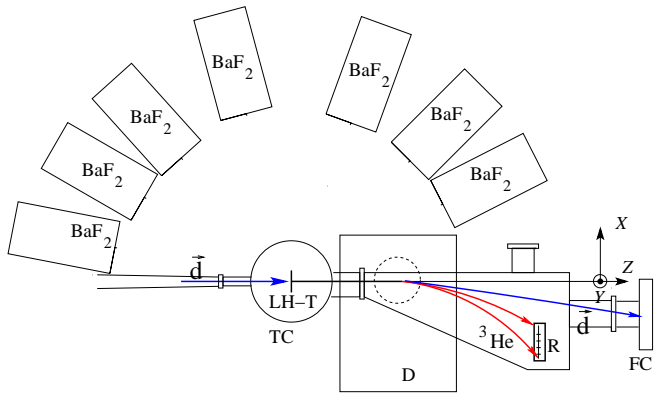}
\caption{\label{fig:cap-setup}
        {(Color online) Schematic overview of the setup: LH-T - liquid hydrogen
	target, D - dipole magnet to separate  $^{3}$He and unscattered
	deuterons, R - recoil-detectors, 
	 FC - Faraday cup.}}
\end{figure}

Figure \ref {fig:cap-setup} gives a detailed view of the setup for the measurements of the 
$\mathrm{^1H(\vec{d},\gamma)^3He}$--capture reaction. The polarized
deuteron beam  was incident on a liquid hydrogen target (LH-T) with a thickness of 14~mg/cm$^2$ enclosed 
by 2.5~$\mathrm{\mu m}$ Havar windows. The target cell was cooled to about 16~K with a closed--cycle
helium--refrigerator and operated at a pressure of 0.25~bar. It was mounted in a specially
designed scattering chamber with 3~mm Al--walls and a conical entrance beam tube to allow for 
measurements at very large scattering angles.

The capture photons were detected by
four large BaF$_2$--counters in coincidence with the recoil $^3$He particles detected in thin
plastic scintillators (R). To separate the beam from the recoil particles a C--shaped dipole--magnet (D) with a vertical pole-tip distance of 90~mm was used downstream of the interaction point.

\subsubsection{$\gamma$-detectors} \label{sec:g-detectors}

The photons from the capture reaction with energies from 13~MeV to 16~MeV for $E_{d}=$29~MeV and from 17~MeV to 24~MeV for 
$E_{d}=$45~MeV were detected with four BaF$_{2}$--scintillators. The detectors were placed at a distance
of 80~cm from the target at various angles in the range from 27$^{o}$ to 169$^{o}$. Each detector consists of four large
cubic crystals \mbox{8$\times$8$\times$25 cm$^{3}$} placed in aluminium containers with a wall-thickness of 5~mm. The boxes 
were shielded with 5~cm of lead on the sides and 5~cm borated plastic in front. 

The scintillator material was chosen because of its excellent timing characteristics. The light response of BaF$_{2}$
is characterized by two decay times of \mbox{0.7~ns} (short) and
\mbox{620~ns} (long) and a light emission spectrum with maxima at \mbox{220 nm} and  \mbox{310 nm}, respectively \cite{Laval83}.
The BaF$_{2}$ fast component allows for an excellent discrimination between the capture photons and the copious
neutrons from break--up reactions and beam--target window interactions. This is particularly important for the 
measurements at extreme angles where the capture cross section is very small.
Due to its high density of \mbox{4.89 g/cm$^{3}$} BaF$_{2}$ has a very good efficiency and 
an acceptable energy resolution of about 16\% (short component) for $E_{\gamma}$$\approx$ 20~MeV. 
Each crystal is connected to fast photomultiplier tubes Philips XP4318B with special optical gel transparent for 
UV--radiation (\mbox{\it {Baysilone~\"Ol}} M~600000 by GE Bayer Silicons GmbH).
A light emitting diode, operated at a frequency of 10~Hz, allowed to monitor the gain during the experiment. 
At forward angles the standard $\mu$-metal shielding of the photomultiplier tubes has been complemented with two closed 
$\mu$-metal boxes with a wall-thickness of \mbox{1.5 mm}, shielding the whole detectors from the magnetic field of the 
C--magnet of about \mbox{50~G}.

\subsubsection{Recoil-detectors}  \label{sec:r-detectors}

To detect the  deflected recoil particles with energies of 17--21 (26--31)~MeV at the beam energy of 29~MeV (45~MeV) plates of plastic scintillator {\it{Pilot U}} with a thickness of 1.2~mm were used. The $^3$He recoiled in a cone between 0.4$^{o}$ and 2.6$^{o}$ at both deuteron energies depending on the photon angle. The strength of the magnetic field was chosen to deflect the initial deuteron beam by $\sim$ 8$^{o}$ such as to separate  the $^{3}$He and deuterons by more than 10$^{o}$. 

The position and size of the recoil detectors were defined by numerical simulations of deuteron and $^{3}$He trajectories in a magnetic field based on measurements of the magnetic field map determined at PSI. Angular and spatial deviations due to beam divergence, beam size and multiple scattering in the target was taken into account. The  detectors were designed to accept 99\% of the recoil particles. The size of the individual {\it{Pilot U}}-pieces was chosen to yield nearly equal $^{3}$He fluxes in all detectors to protect the electronics from signal over-load. The thickness was chosen by the requirement to stop the recoil $^3$He but not the Rutherford scattered deuterons and the protons from break--up reactions. The photomultipliers were mounted vertically at a distance of about 50~cm where the magnetic field from the deflection magnet could be sufficiently shielded.

\subsubsection{Electronics} \label{sec:Electronics}

All the detector signals were multiplexed to form a trigger signal and to process the signal for time and amplitude
measurements. The trigger signals were clipped to correct for base line shifts at high rates and fed to constant
fraction discriminators to minimize the signal amplitude dependence. A coincidence was requested between each recoil
and BaF$_2$ detector within a time window of 25~ns. The sum of all coincident signals was used for a further coincidence
with the RF--signal of the cyclotron. To form these coincidences first is important to minimize electronic dead time
effects at the high rates. The final trigger signal with the RF time provided the start of the TDC and triggered the read--out of the CAMAC system. For each detector the retimed coincidence signal was used for the gate of the charge integrating ADC. For redundancy
coincidences between the recoil detectors and the BaF$_2$ detector were also formed. For the BaF$_2$ 
detectors two amplitudes were recorded. One with a short 25~ns gate for the fast component and one with a 700~ns gate 
for the long one. For a time of flight (TOF) measurement, the TDC's were started with the beam RF signal and stopped
with the individual fast component detector signal. The same was done for the recoil detector signals. 
A signal from the digitized beam current of the Faraday cup, a real-time clock and pulser signals as well as the current polarization 
state signal were fed into scalers. These signals were relevant to correct false asymmetries from
dead time and beam current variations correlated with the polarization state. All the information was written on an 
event--by--event basis on disc for on--line analysis and also written to tape for backup and replay.

\section{Analysis and Results}
\begin{figure}[htb]
\includegraphics[width=0.49\textwidth,clip]{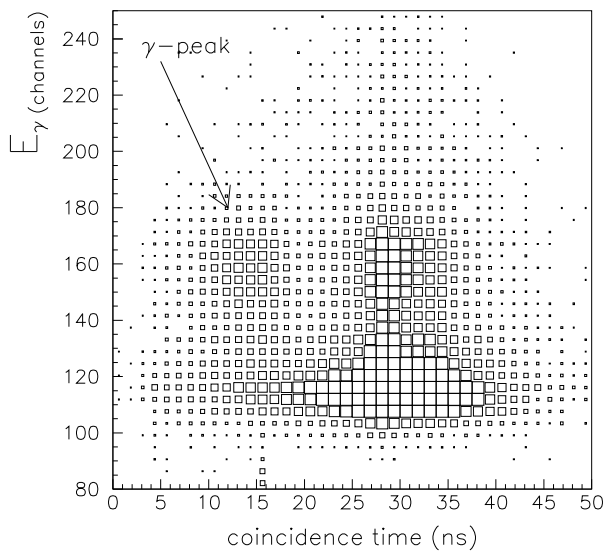}  
\caption{Example of a spectrum of the short component amplitude of the BaF$_{2}$ response 
versus the TOF between the $\gamma$-rays and the recoil particles. Cuts are applied on the
recoil energy and on the BaF$_{2}$--RF--TOF.
\label{fig:coin}}
\end{figure}
The main challenge of the data analysis is to single out the few capture events from a huge background due to hadronic
reactions. Particularly for data taken at the very large and very small scattering angles the applied hardware coincidence 
is indispensable. The coincidence requirement results in a signal to noise ratio of the order of 1 for these data. The
remaining background is mostly due to accidental coincidences and $n-p$ breakup reactions of deuterons on hydrogen and 
the target windows. Depending on the kinematics the signal to noise ratio could be enhanced to at least 25 due to
the excellent timing resolution of the BaF$_2$ detectors and the efficient software time cuts on the prompt $\gamma$ events.
In order to achieve this result software cuts have been applied on the BaF$_{2}$--recoil--TOF, the 
BaF$_{2}$--RF--TOF, and on the energy information of the recoil--detectors. In addition, due to the kinematical 
angular correlation the proper recoil detector is selected for a given BaF$_{2}$. This selection provides for a 
significant reduction of background.

\begin{figure}[htb]
\includegraphics[width=0.49\textwidth]{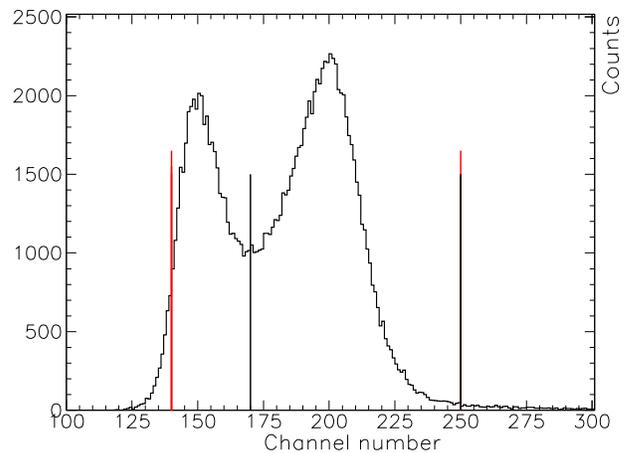}  
\caption{(Color online) Example of a $\gamma$--energy spectrum with all cuts applied and the integration limits for the capture events.
For studies of the background analyzing power as discussed in the text, a third integration line including the background is also shown\label{fig:short}}
\end{figure}

Figure~\ref{fig:coin} shows an example of a two-dimensional histogram with a BaF$_{2}$
detector lightoutput versus its BaF$_{2}$--recoil--TOF of the corresponding recoil detector. The cuts on the BaF$_{2}$--RF--TOF
and recoil lightoutput are applied. The two-dimensional histogram 
shows a clear separation between the $\gamma$$-$$^{3}$He coincidences and the unstructured 
accidental ones. With an additional cut on the coincidence time BaF$_{2}$--energy spectra like the one shown in
figure \ref{fig:short} have been achieved.

In  figure~\ref{fig:short} a $\gamma$--energy spectrum (BaF$_{2}$ short component) with all mentioned cuts applied   
(except the cut on BaF$_{2}$-lightoutput) is shown. The spectrum also shows the integration limits applied to determine
the capture events. The left part of the spectrum is due to the remaining low energy $\gamma$ 
background which also contributes to the region of the $\gamma$-peak within the integration limits.
To determine the effect of this remaining background contribution the response function of the BaF$_{2}$--detector to monoenergetic 
photons must be known. 

\begin{figure*}[htb]
\includegraphics[width=0.49\textwidth]{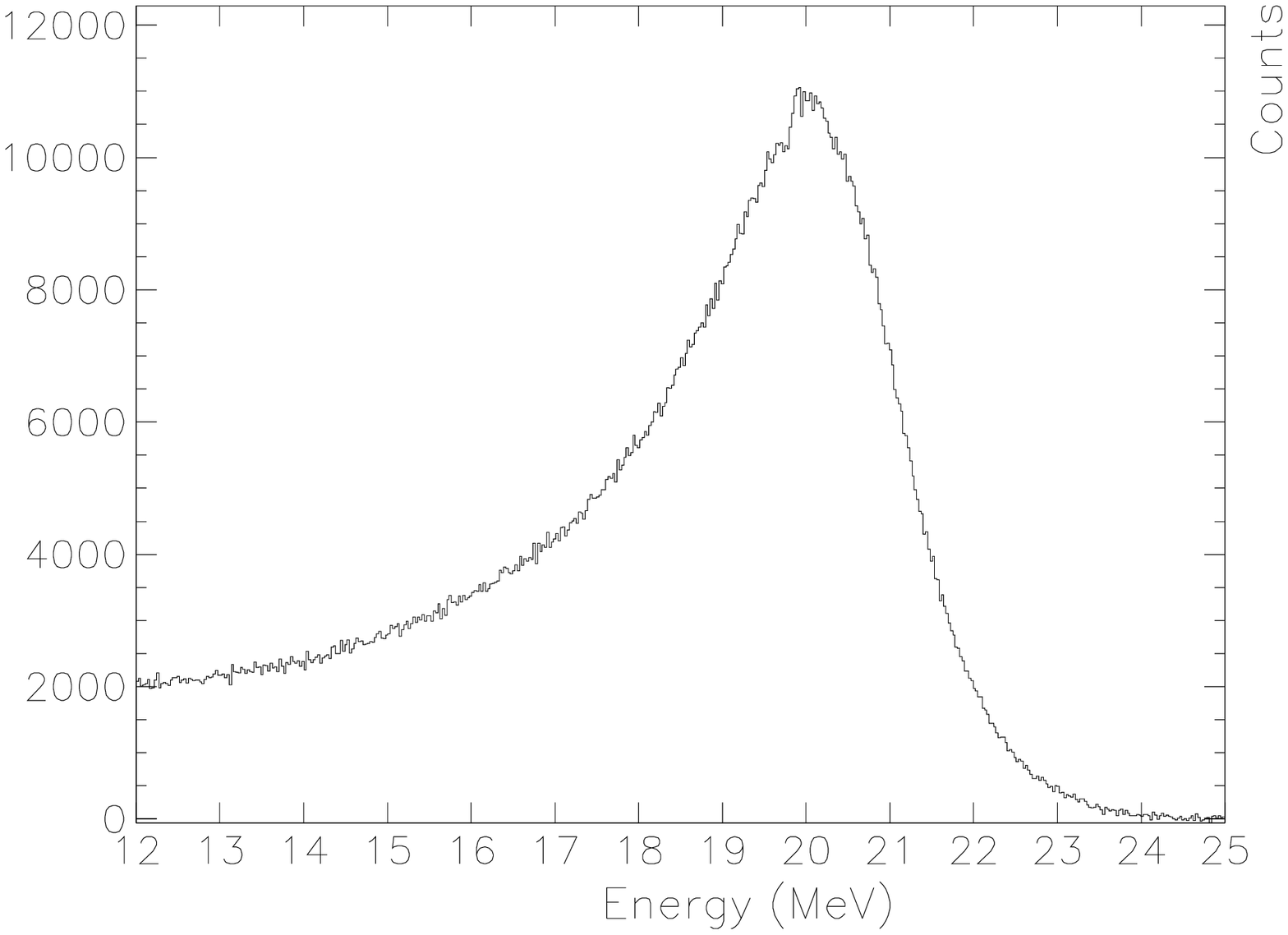}\includegraphics[width=0.49\textwidth]{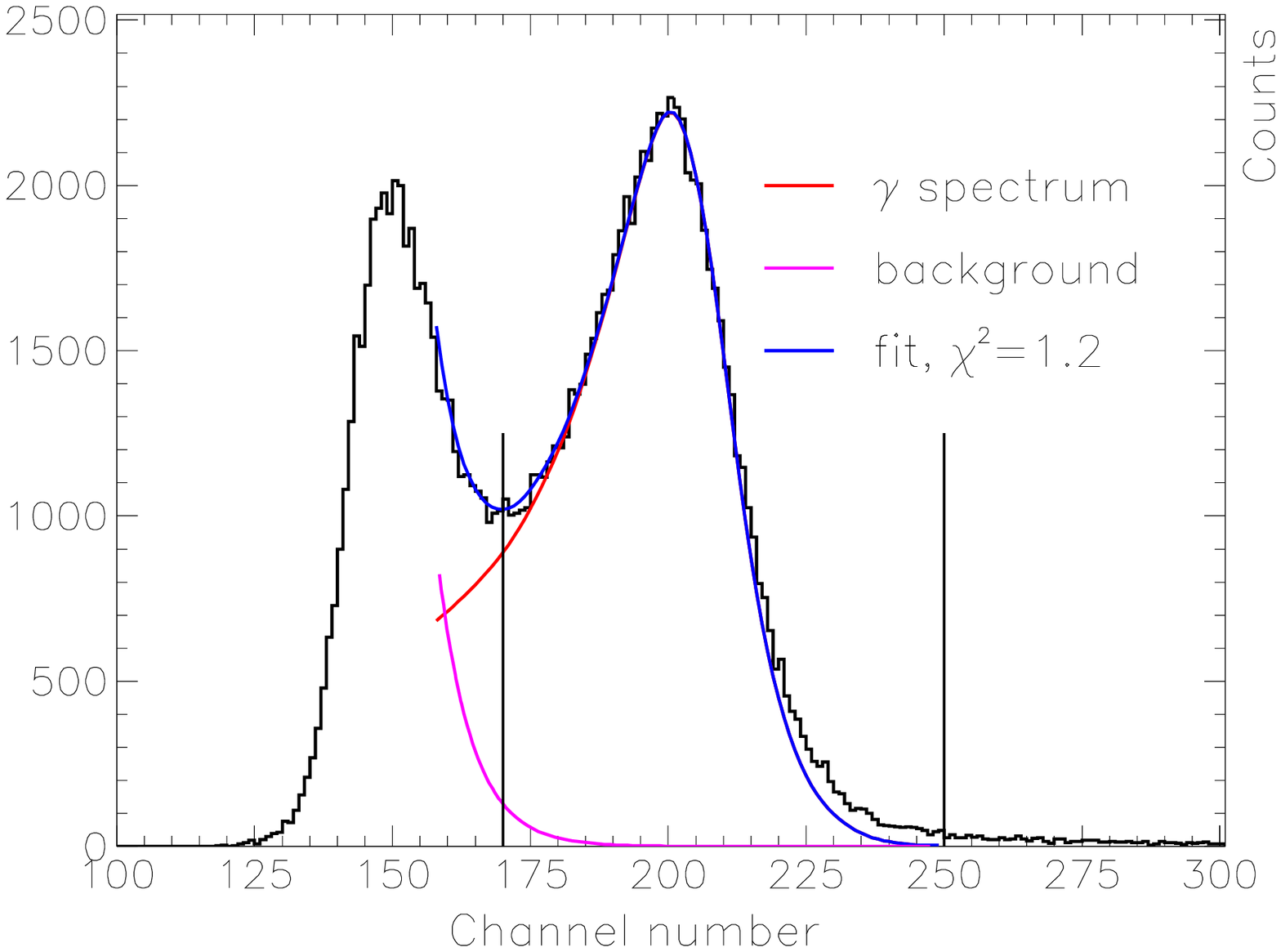}   
\caption{(Color onlie) On the left the short component of the BaF$_{2}$ response function for 20~MeV $\gamma$--rays.
On the right experimental $\gamma$-spectrum in comparison with the fit-function. The integration limits are represented by two vertical lines. The contribution of the exponential function and the folded peak
function are also shown separately
 \label{fig:both}}
\end{figure*} 
For this determination an additional experiment with monoenergetic $\gamma$-rays was performed at the Physics Institute of the University of Basel. Monoenergetic 20 MeV $\gamma$-rays were produced in a $^3$H(p,$\gamma$)$^4$He 
reaction with a 1~MeV proton beam provided by a Cockroft-Walton accelerator. The detector was placed at 110 deg close to the maximum of the angular distribution 
of the photons. The distance between the tritium target and the BaF$_{2}$-detector 
was 80 cm in order to reproduce 
the geometry of the $\vec{d}-p$-capture experiment. The target was a 0.45 mg/cm$^{2}$ Ti layer on a Cu--plate with 2~Ci absorbed 
tritium. The electronics was a simplified version of the $\vec{d}-p$-capture experiment using the same electronic 
modules and gate widths. The measured short component of the BaF$_{2}$ response function 
for the monoenergetic photons from this study is shown on the left in figure ~\ref{fig:both}.

The sum of the experimentally determined response function and an exponential function folded with a Gaussian is used in the analysis to fit the spectra of the short component of the BaF$_{2}$ detectors. The peak position, the amplitude, the
width of the Gaussian and the slope of the exponential function are used as parameters. An example of such an analysis is 
shown on the right in figure~\ref{fig:both}. The reduced $\chi^{2}$ of the fits varies from 0.8 to 2.2. The upper end of the peak is distorted due to pile--up during the $\vec{d}-p$--capture experiment and thus is excluded in the fit.

As the energy of the photons in the $\vec{d}-p$-capture experiment varied between 13~MeV and 24~MeV the energy dependence of the response function was studied with a GEANT-Monte-Carlo simulation. Photons with
energies of 13~MeV or with 24~MeV incident on a BaF$_{2}$ crystal folded with a Gaussian to fit the experimental peak width have been 
compared  to a simulation at 20~MeV, the energy of the model peak. Scaling the response functions to the model peak energy resulted in an energy dependence of the response of less than 5\% for the integration limits applied in the analysis. This results in a negligible relative error contribution of 0.2\% for a background contribution of 4.6\%.

With the assumption of an unpolarized background the contribution from 0\% to 4.6\% within the integration limits
can be interpreted as a dilution factor $N^{tot}/(N^{tot}-N^{backgr})$. To account for this dilution
the corrected analyzing powers are then given as
\begin{eqnarray} 
A_{y,~yy}^{capture} =  A_{y,~yy}^{extracted} \cdot  \frac{N^{tot}}{N^{tot}-N^{backgr}} \label{corr}
\end{eqnarray}                                                                                
Here $N^{tot}$ is the total number of events within the integration limits and $N^{backgr}$ is the 
total number of counts of the fitted exponential function within the integration limits.
A$_{y,~yy}^{extracted}$ are the vector--, tensor--analyzing powers determined from the total number of counts in the different 
polarization states, respectively. 

In order to verify the assumption of an unpolarized background the following tests have been performed. First, the
corrected analyzing powers at a given kinematic point must be independent of the dilution factor. Different sets
of runs with different background conditions due to different beam currents have been compared. Within the statistical 
errors no systematic deviations have been found.
Second, if the analyzing powers of the background would not be zero, they would contribute to the extracted analyzing power
values as:
\begin{eqnarray} 
A_{y,~yy}^{capture} =&&  A_{y,~yy}^{extracted} \cdot \frac{N^{tot}}{N^{tot}-N^{backgr}}\nonumber\\ 
&&- A_{y,~yy}^{back} \cdot \frac{N^{back}}{N^{tot}-N^{backgr}}
\label{ayyback}
\end{eqnarray}
Large background contributions with finite analyzing powers would significantly distort the corrected capture values.
This can be checked by artificially increasing integration limits in order to include a large background fraction
in the low-energy part of the spectrum (see figure \ref{fig:short}). The corrected analyzing powers from these integration limits can be compared
to analyzing powers within integration limits with essentially no background contribution. It is found that the 
analyzing powers calculated with the larger integration limits are completely consistent with the analyzing powers 
calculated for the capture peak. This confirms that within the statistical limits
the background contribution is unpolarized, hereby justifying the correction described with equation \ref {corr}.

The resulting deuteron vector (A$_{y}^d$) and tensor (A$_{yy}$) analyzing powers from the capture reaction
are given in table \ref{tab:res29} and \ref{tab:res45}. The statistical and the systematic errors are listed separately. As can be seen the statistical errors dominate the total error for all data points measured.

\begin{table}[htb]
\begin{ruledtabular}
\begin{tabular}{rrr}
$\theta_{c.m. (d-\gamma)}$ & $A_{y}\pm \delta A_{y}^{stat} \pm \delta A_{y}^{sys}$ & $A_{yy}\pm \delta A_{yy}^{stat} \pm \delta A_{yy}^{stat}$ \\
deg  & $\times 100$ & $\times 100$  \\ [1mm]
\hline
%& & & \\[-4mm]
33.52  &$  -3.821 \pm   0.335 \pm  0.084  $&$  3.010 \pm   0.383 \pm   0.090 $  \\
55.32  &$  -1.808 \pm   0.252 \pm  0.061  $&$  2.197 \pm   0.288 \pm   0.070 $  \\
76.41  &$  -0.443 \pm   0.262 \pm  0.050  $&$  2.236 \pm   0.300 \pm   0.053 $  \\
116.16 &$   0.465 \pm   0.253 \pm  0.051  $&$  2.452 \pm   0.289 \pm   0.054 $  \\
134.94 &$   1.030 \pm   0.285 \pm  0.052  $&$  2.182 \pm   0.326 \pm   0.055 $  \\
153.19 &$   0.325 \pm   0.470 \pm  0.044  $&$  0.438 \pm   0.538 \pm   0.045 $  \\
170.21 &$   1.097 \pm   0.412 \pm  0.060  $&$ -2.934 \pm   0.520 \pm   0.067 $  \\
\end{tabular}
\end{ruledtabular}
\caption{Vector (A$_{y}^d$) and tensor (A$_{yy}$) analyzing powers for the $\vec{d}-p$ radiative capture 
reaction at a deuteron energy of 29~MeV.\label{tab:res29}}
\end{table}

\begin{table}[htb]
\begin{ruledtabular}
\begin{tabular}{rrr}
$\theta_{c.m. (d-\gamma)}$ & $A_{y}\pm \delta A_{y}^{stat} \pm \delta A_{y}^{sys}$ & 
                          $A_{yy}\pm \delta A_{yy}^{stat} \pm \delta A_{yy}^{stat}$ \\
deg  & $\times 100$ & $\times 100$  \\ [1mm]
\hline
%& & & \\[-4mm]
31.04  &$  -5.134 \pm   0.186 \pm  0.095  $&$ 4.610  \pm 0.228  \pm  0.113   $ \\
51.20  &$  -1.955 \pm   0.160 \pm  0.042  $&$ 1.199  \pm 0.197  \pm  0.048   $ \\
112.88 &$   1.837 \pm   0.190 \pm  0.042  $&$ 1.891  \pm 0.236  \pm  0.049   $ \\
137.89 &$   2.986 \pm   0.139 \pm  0.057  $&$ 2.014  \pm 0.171  \pm  0.068   $ \\
170.48 &$   4.507 \pm   0.328 \pm  0.081  $&$ 1.334  \pm 0.407  \pm  0.095   $ \\  
\end{tabular}
\end{ruledtabular}
\caption{Vector ($A_{y}^d$) and tensor ($A_{yy}$) analyzing powers for the $\vec{d}-p$ radiative capture 
reaction at a deuteron energy of 45~MeV.\label{tab:res45}}
\end{table}

Various systematic errors are accounted for. A relative 10\% error is estimated for the background contribution.
In addition, correlations between luminosities and polarization states could lead to false asymmetries. To determine a systematic
error the asymmetries of two unpolarized runs with different current have been used to determine a false asymmetry
due to different dead times and luminosities. The determined effect can be scaled to the differences in currents
and dead times for different polarization states present during normal running conditions. A resulting systematic
uncertainty of 0.00043 (0.00024) results due to current differences of 0.005(0.003)~nA during the data taking at 
$E_{d}$=29(45)~MeV. Including the uncertainty of the polarization determination the total systematic errors vary 
between 0.00044 and 0.00113 compared to the statistical errors from 0.00187 to 0.00470.

\section{Comparison to theory}

\begin{figure*}[bht]
\includegraphics[width=0.45\textwidth]{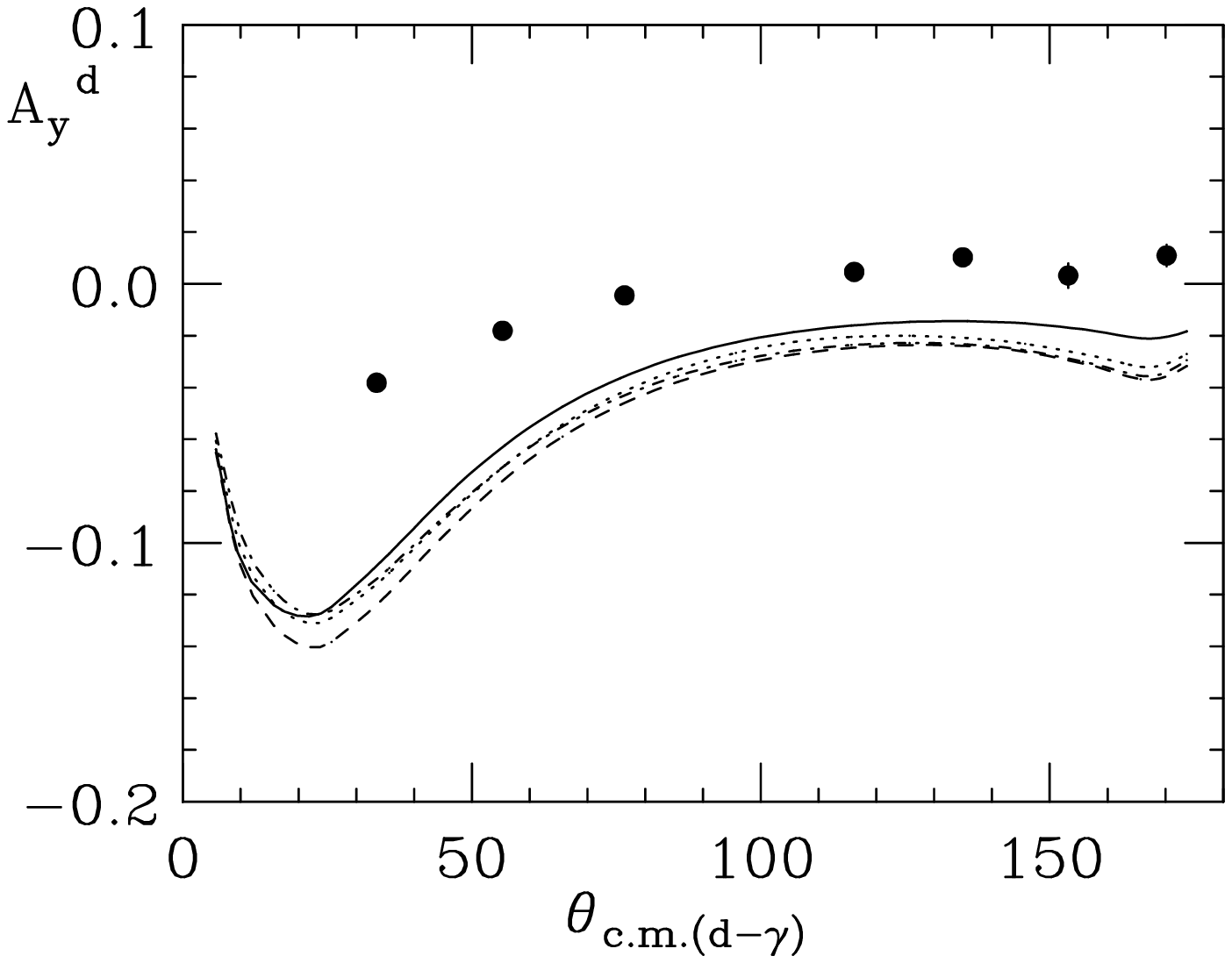}\includegraphics[width=0.45\textwidth]{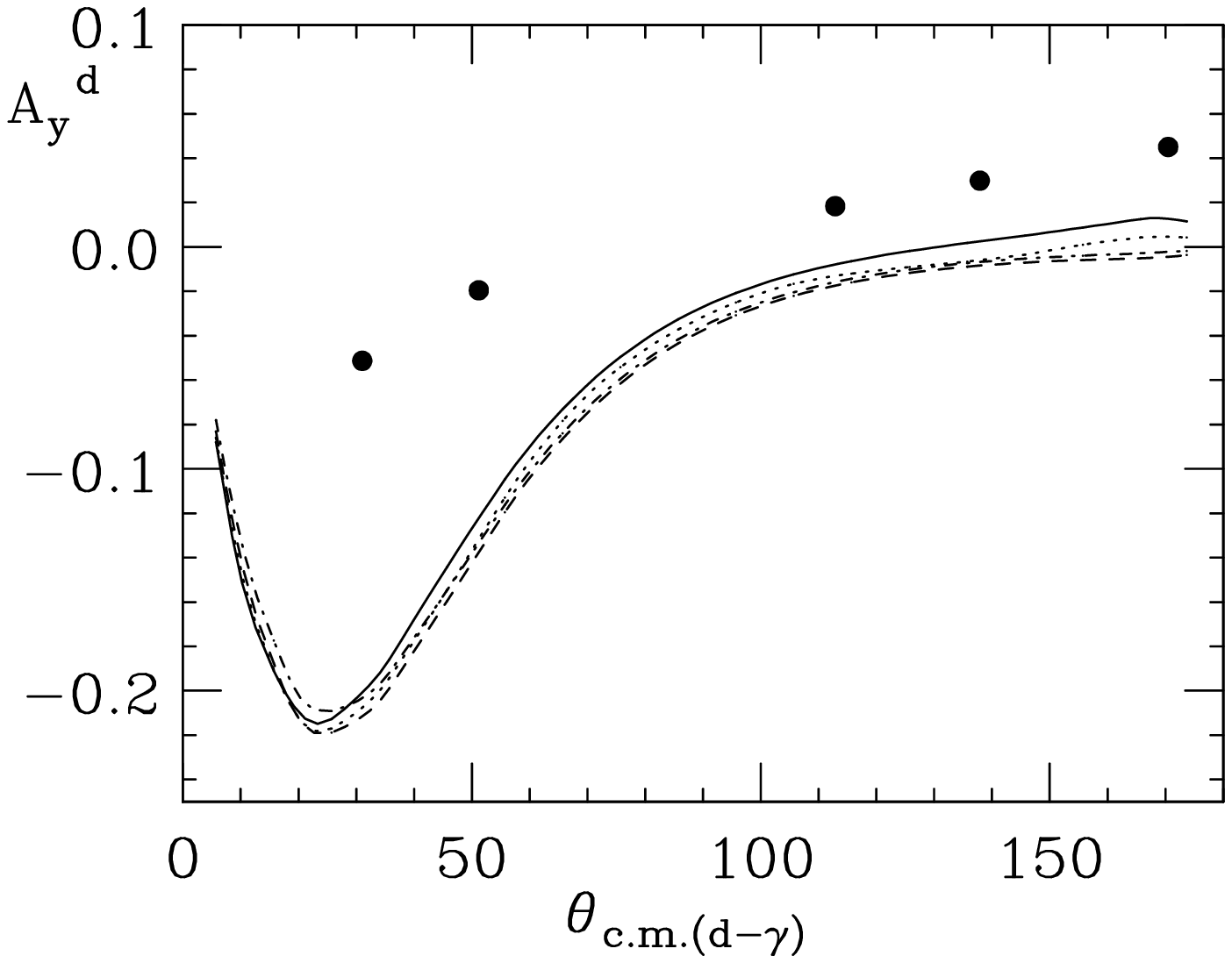}   
\includegraphics[width=0.45\textwidth]{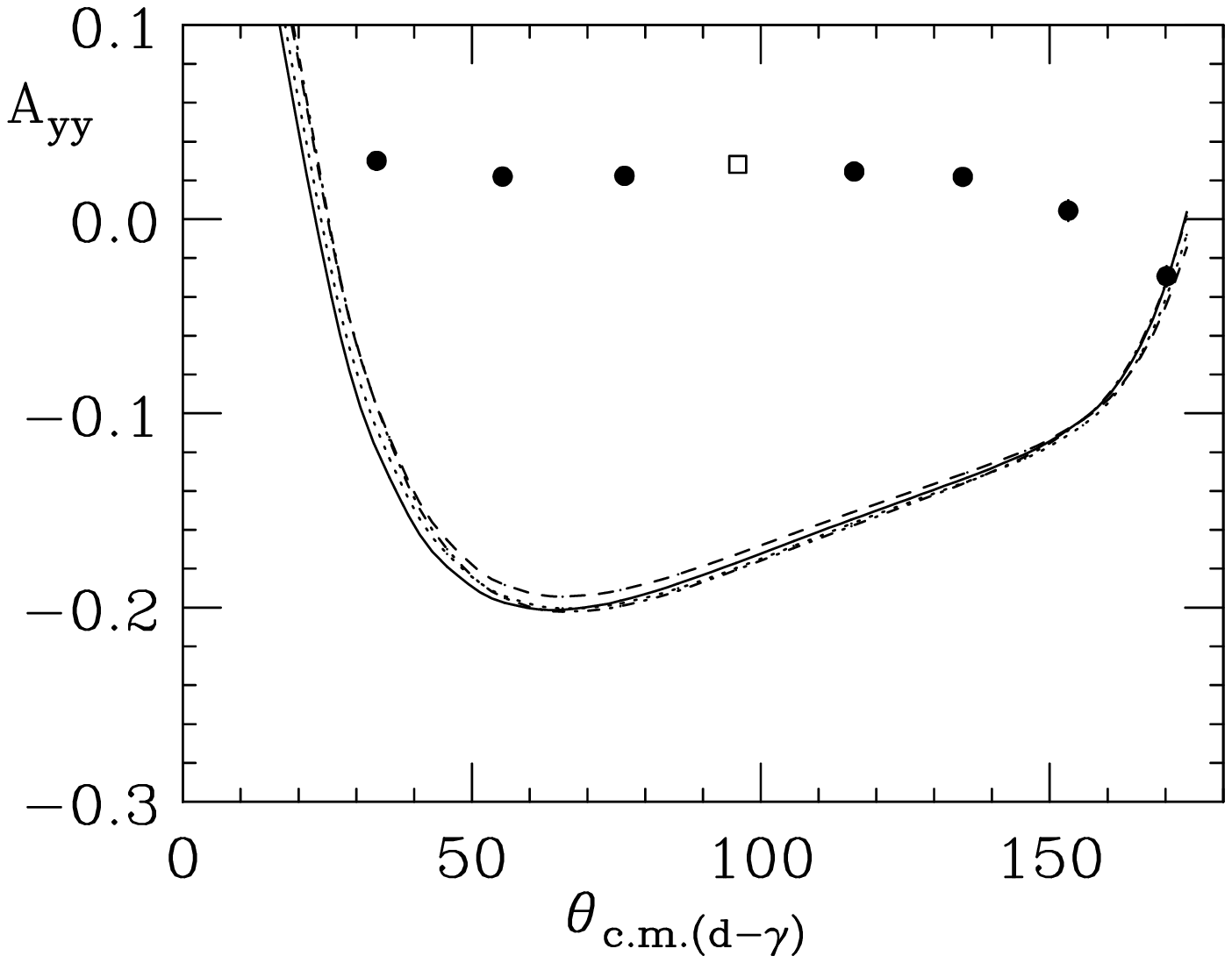}\includegraphics[width=0.45\textwidth]{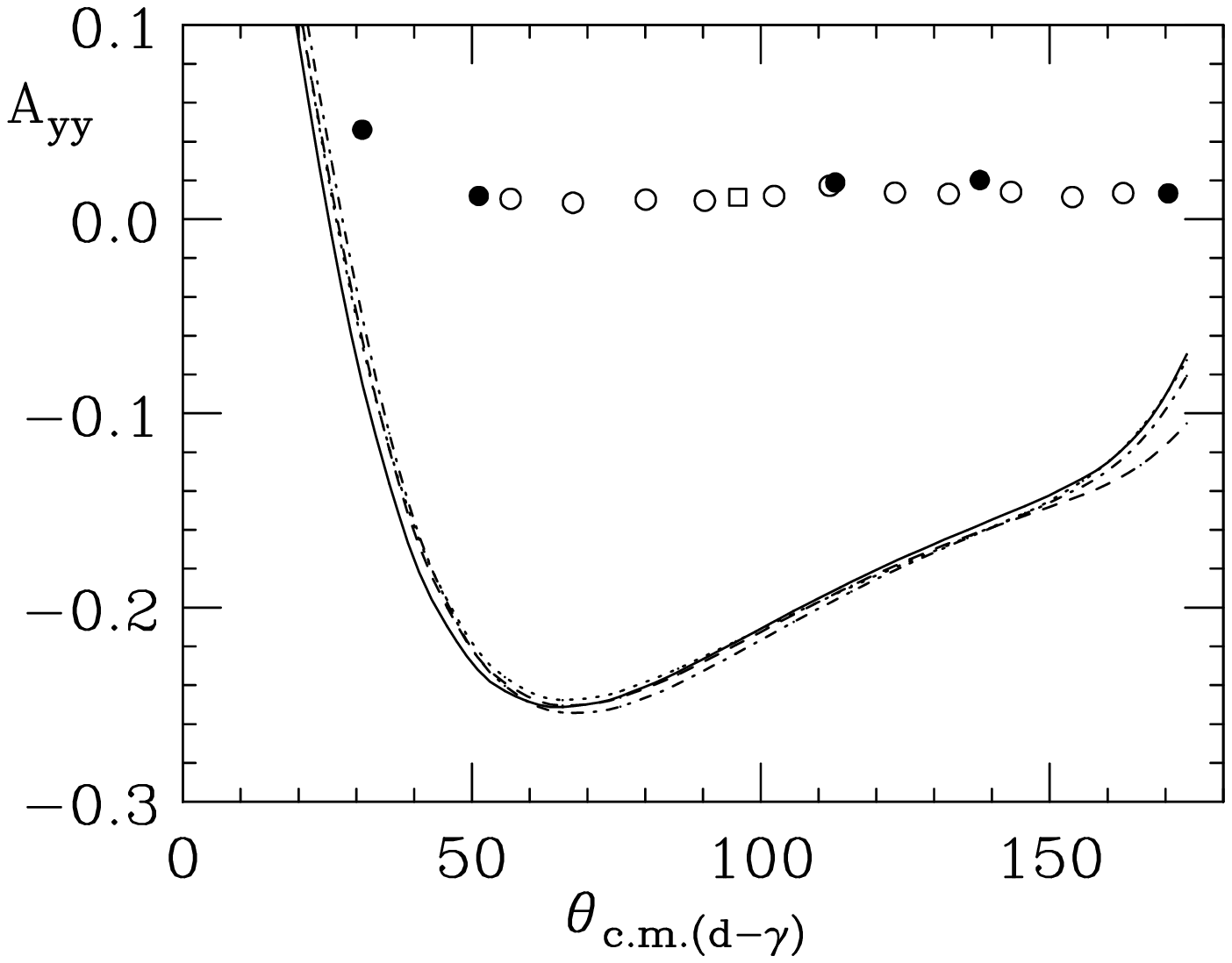}   
\caption{\label{fig:1body} A$_y^d$ (top) and A$_{yy}$ (bottom) for 29~MeV (left) and 45~MeV (right) incident deuteron energy as a function of the center--of--mass angle between deuteron and outgoing $\gamma$. Data of the present experiment ($\bullet$) together with the data
by Anklin et al {\protect{\cite{Anklin98}}} ($\circ$) and Jourdan {\em et al.} {\protect{\cite{Jourdan86}}} ($\square$) are compared to the one--body--calculations by Deltuva {\em et al.} (solid), Skibinski {\em et al.} with/without 3BF (dotdash/dash), and Viviani {\em et al.} (dot).}
\end{figure*} 
The present data together with previous data taken at the same deuteron energies are compared to three different recent calculations \cite{Skibinski03,Marcucci05,Deltuva04b}. The calculations are all
exact in the sense that they provide a full solution of the Schr\"odinger equation from a realistic nucleon--nucleon interaction for both the ground-- and continuum states. An exact treatment using the same Hamiltonian for initial and final state is known to be essential for a successful description of polarization observables of the $\vec{d}-p$ capture reaction. Whereas for the description of the differential cross section the use of the Born approximation leads to an underestimate 
of order 15\%, the tensor analyzing power A$_{yy}$ is underpredicted by more than a factor of 2. The importance of 
initial state interaction in these observables has been observed already in the calculation by J.~Torre \cite{Jourdan86}.
More recently a calculation by A.C.~Fonseca and D.R.~Lehman \cite{Fonseca91} confirmed that re-scattering effects in the 
initial state are crucial for a determination of polarization observables.

The techniques applied in the calculations by the three groups are very different.
The Bochum--Cracow group \cite{Skibinski03} obtains the wave functions by solving the Faddeev equations in a non--relativistic framework. 
In the calculation of the PISA group \cite{Marcucci05} the wave functions are calculated using the pair--correlated hyperspherical harmonics method \cite{Kievsky01}. Both groups use the Argonne v18 two--body potential \cite{Wiringa95} as underlying nucleon--nucleon potential. In addition, the Urbana IX three--body potential \cite{Pudliner95} is also included by both groups. In the approach by the Hannover group \cite{Deltuva04b} the three--particle scattering equations are calculated exactly with a Chebyshev expansion of the two--baryon transition matrix. This approach employs the CD--Bonn potential \cite{Machleidt01} as the underlying two--body--potential. The three--nucleon--force (3BF) is generated here via a coupled channel extension of CD--Bonn which allows for a single nucleon transition to a static $\Delta$ isobar. The CD--Bonn + $\Delta$ extension is as exact as CD--Bonn as it is also fitted to the experimental two--nucleon data up to 350~MeV \cite{Deltuva03a}.

In figure \ref {fig:1body} the results of the three calculations for the analyzing powers at the energies of the experimental data are compared. In this comparison only one--body--currents are included in the calculations. In addition, the results of one calculation by Skibinsky {\em et al.} (dash) is shown which neglects the 3BF. The figures confirm that when accounting for one--body--currents
only, the results are essentially the same. Small deviations can be observed for the calculation which does not include a 3BF (dash) and for the calculation which includes the 3BF as a static $\Delta$ isobar (solid). 
Thus, the effects of the 3BF are small but notable in A$_y^d$ when a different 3BF--model is employed.

Figure \ref {fig:1body} also shows the experimental results of the present work together with the results of \cite{Anklin98} and \cite{Jourdan86}. Whereas the results of the three experiments are in good agreement the comparison to the calculations shows a large discrepancy. As will be shown, the dominant cause of these deviations are the missing two--body--currents in the calculations. One should note that all the theoretical results have been folded with the acceptances of the experimental data
of $\pm 5.7$ deg.

In calculations of capture-- and photodisintegration observables the dominant part of many--body--currents is usually included  using the Siegert theorem \cite{Siegert37}. The standard Siegert theorem is formulated in a multipole expansion in which part of the  transition currents can be replaced by the charge operator. Alternatively is has been shown by Golak {\em et al.} \cite{Golak00a} that in momentum space a multipole expansion of the current operator is not necessary; this approach is applied in the calculations by Skibinski {\em et al.} Whereas two body currents are implicitly included in the dominant part of the electric transitions, they are not accounted for in the magnetic transitions and a small part of the electric ones. In the calculation by Skibinski {\em et al.} only the one--body--current is used in these latter parts of the transition.
An alternative approach is used in which the $\pi$-- and $\rho$--exchange currents are taken into account explicitly using the Riska prescription \cite{Riska89}. 
\begin{figure*}[htb]
\includegraphics[width=0.45\textwidth]{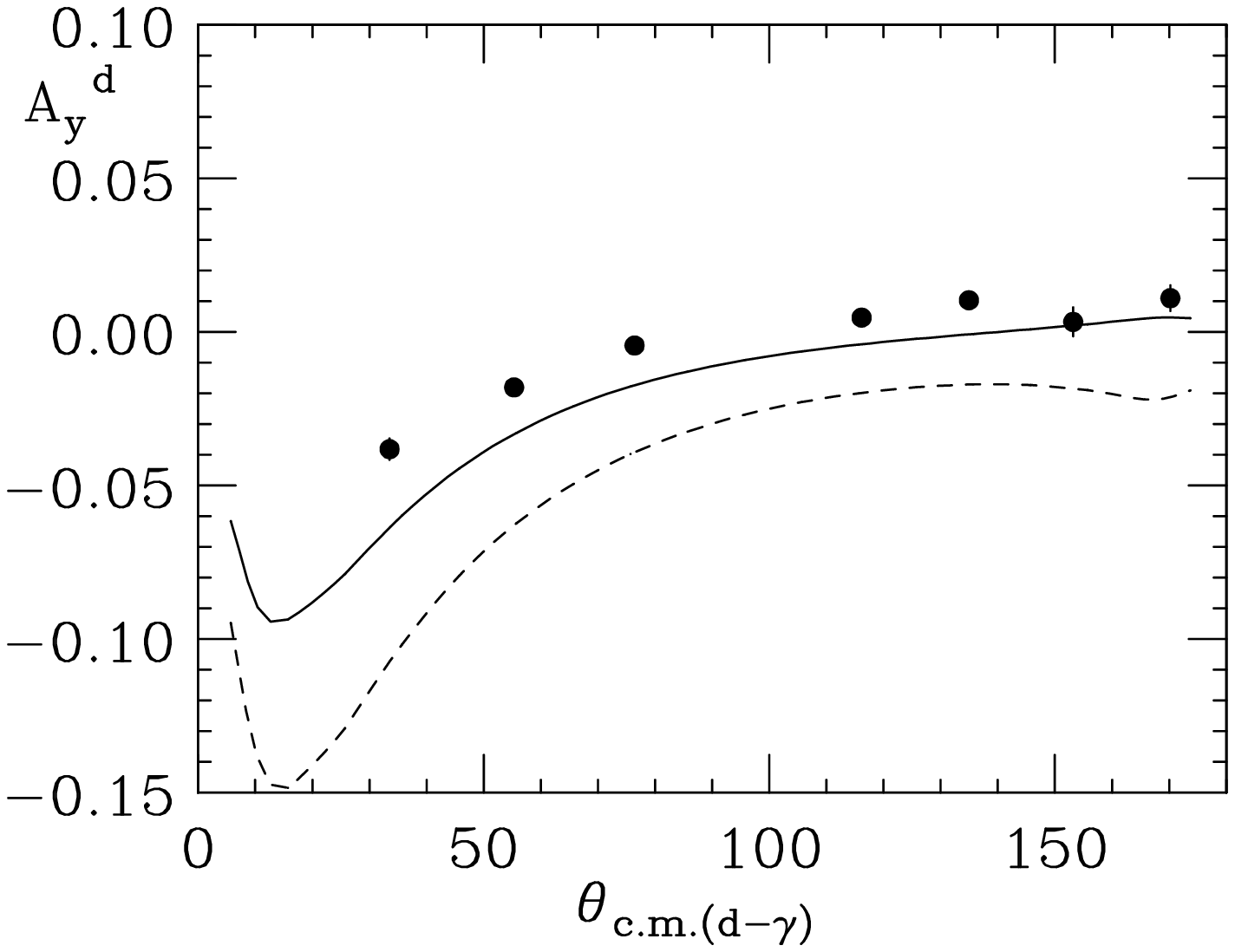}\includegraphics[width=0.45\textwidth]{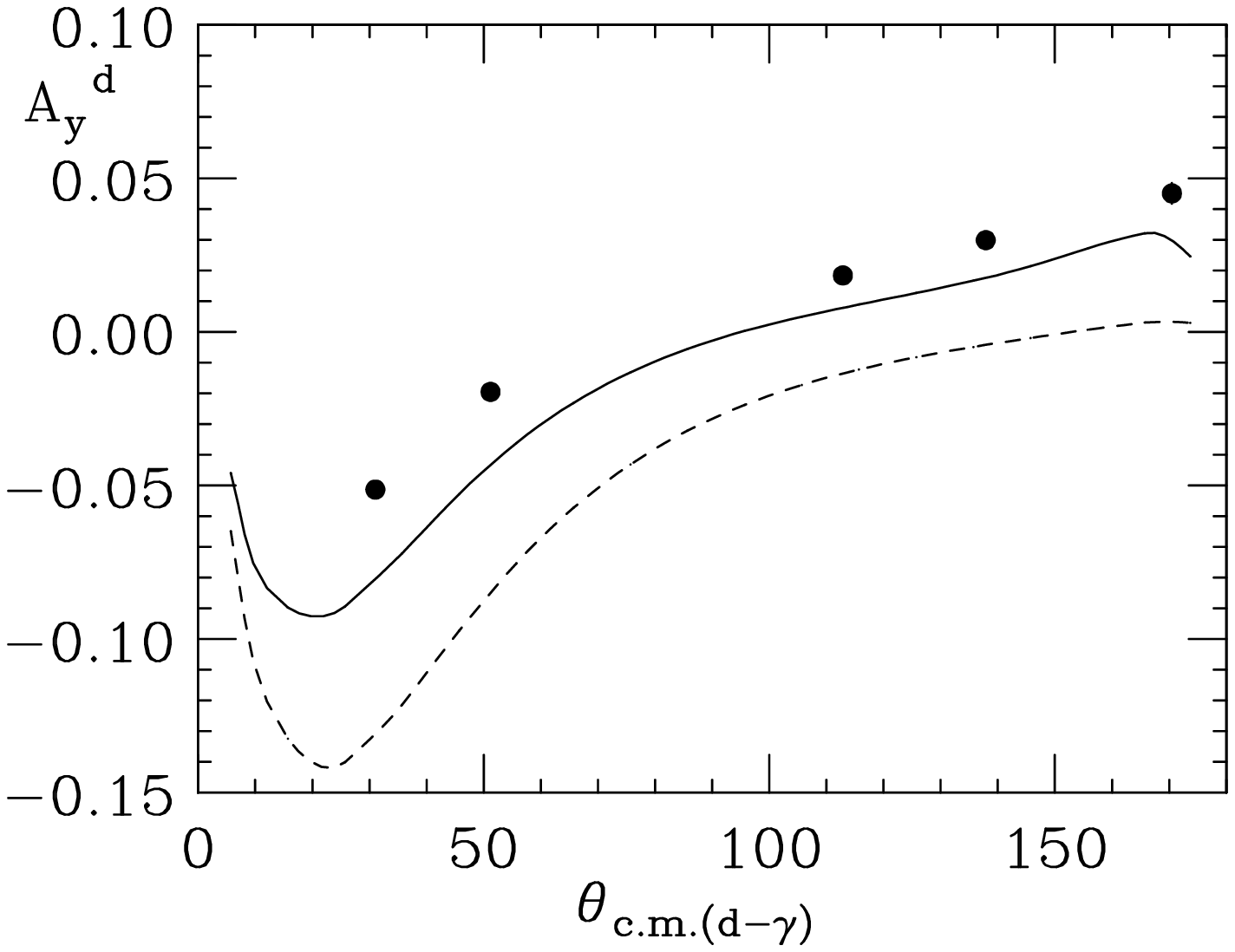}   
\includegraphics[width=0.45\textwidth]{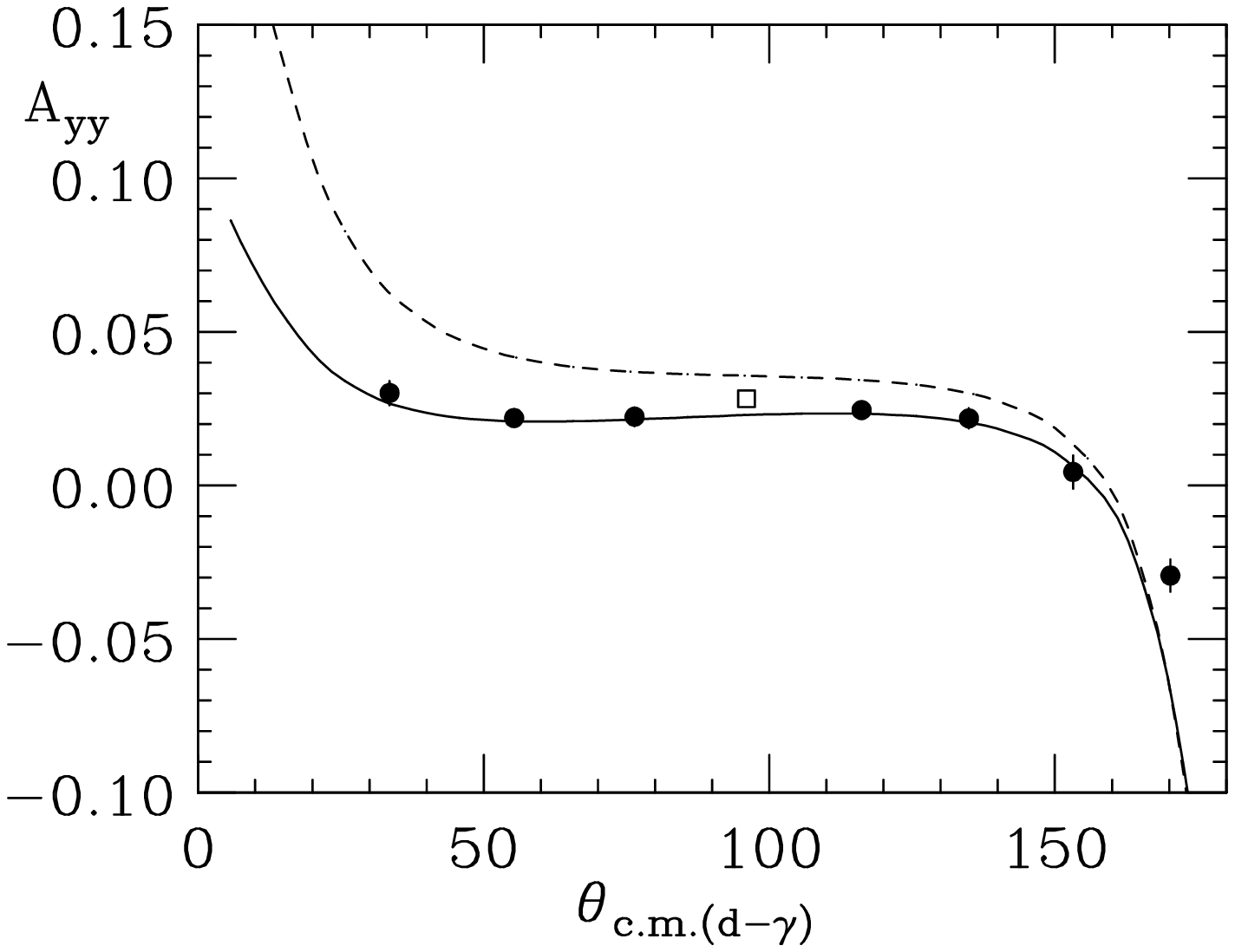}\includegraphics[width=0.45\textwidth]{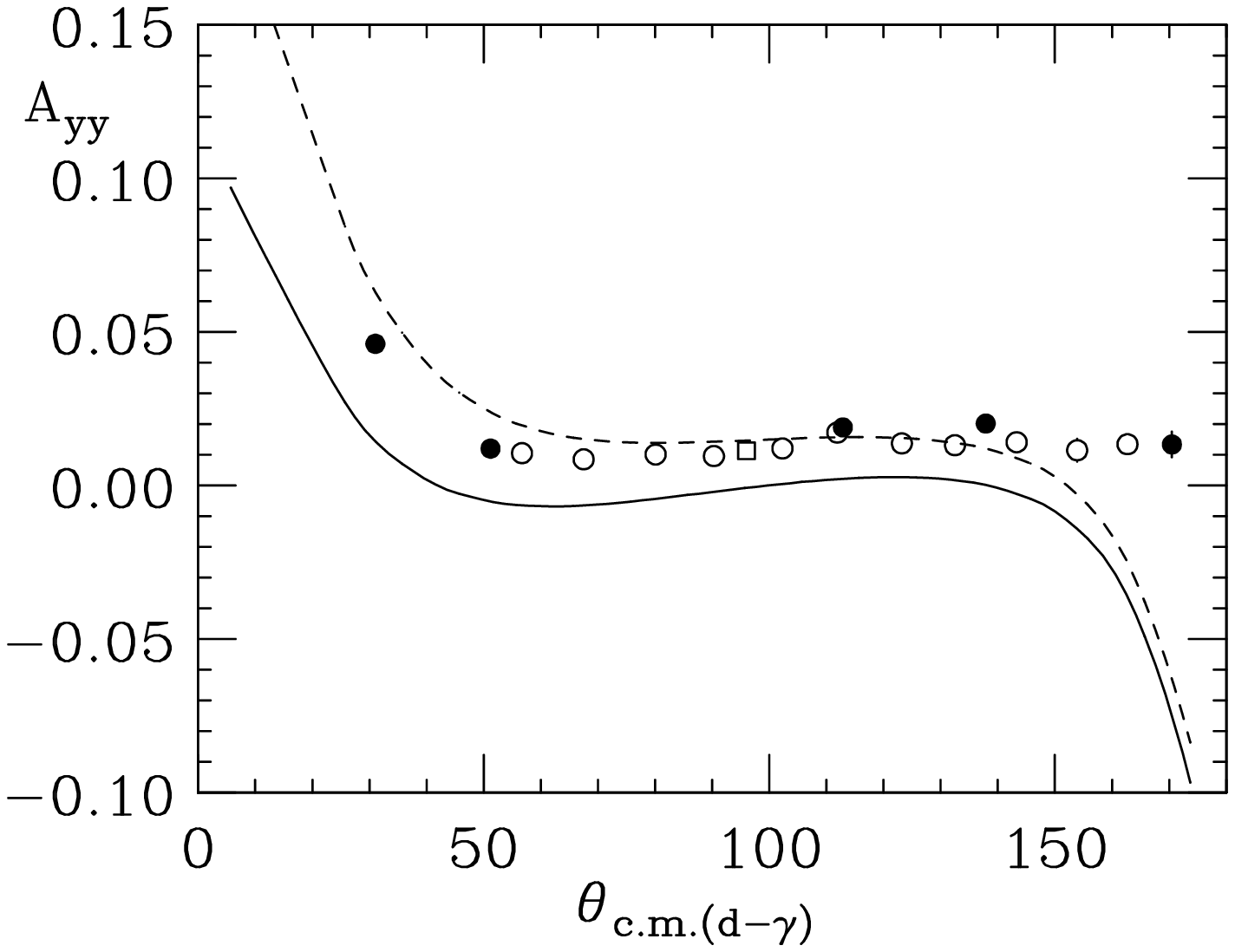}   
\caption{Same as figure {\protect{\ref{fig:1body}}}. Here the data are compared to the calculations by Skibinski {\em et al.} with an explicit treatment of the exchange currents (solid) and within the Siegert approach (dash).  
\label{fig:2body}}
\end{figure*} 

The effect of the two--body--currents is shown in figure \ref{fig:2body}. The description of the data has largely improved which confirms the presence of large two--body--current effects. The solid lines give the results including the $\pi$-- and $\rho$--exchange currents explicitly
whereas dashed lines show the results from the Siegert approach. In general, the description of the data is much better with the explicit treatment of the exchange currents.  This suggests that magnetic transitions,
for which MEC's are not accounted for in the Siegert approach, are a relevant part of the transition. As expected, this holds particularly for the wings of the A$_{yy}$ data. In the intermediate angular range, which is dominated by the E1--transition, one would expect the Siegert calculation, which implicitly accounts for MEC's, to be the more successful approach. However, only the A$_{yy}$ data at the 45~MeV are described well. 
This is surprising as these data are particularly sensitive to small ingredients of the calculations. Thus,
one should conclude that only a more complete explicit account of the two--body--currents can improve the description of the data.
\begin{figure*}[htb]
\includegraphics[width=0.45\textwidth]{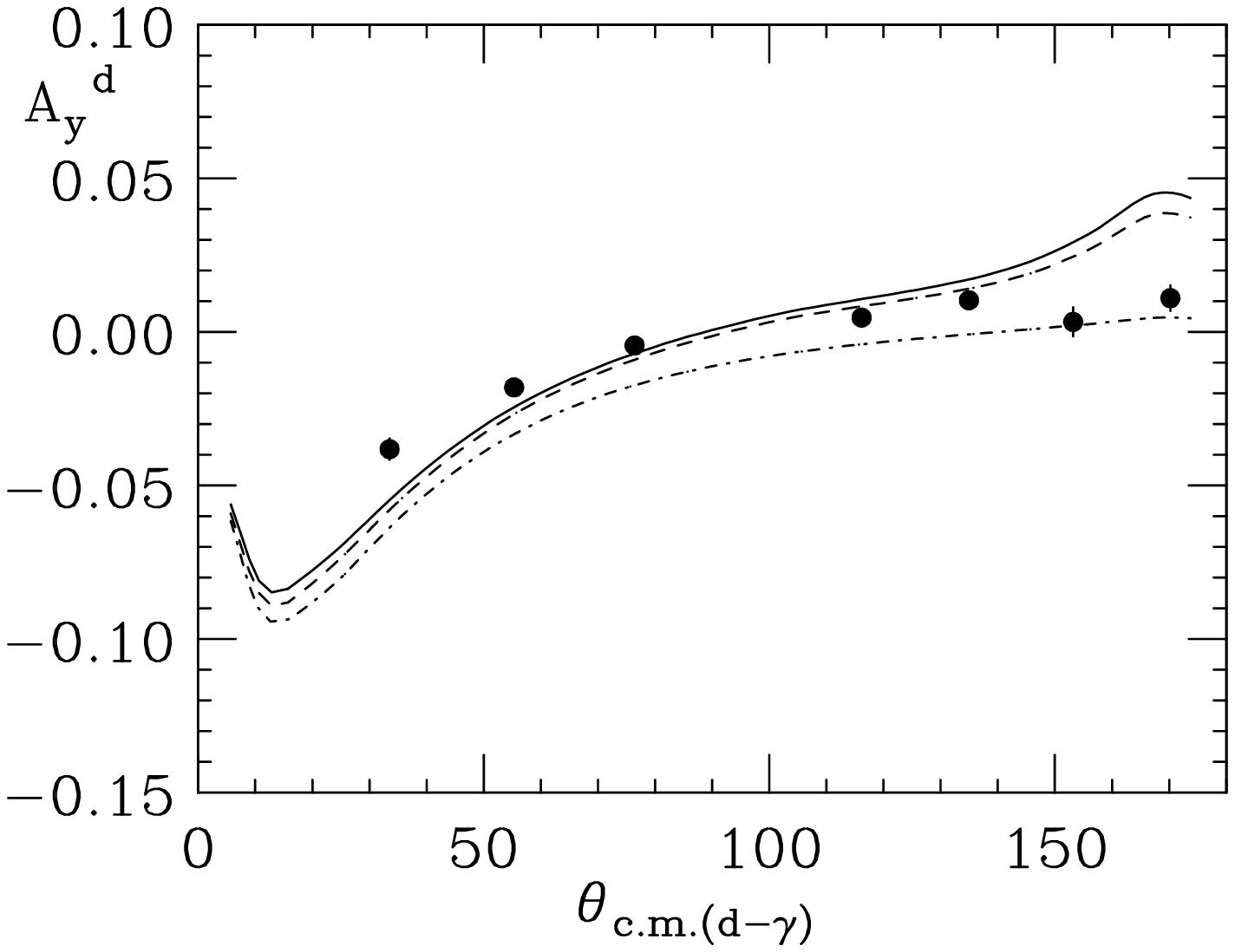}\includegraphics[width=0.45\textwidth]{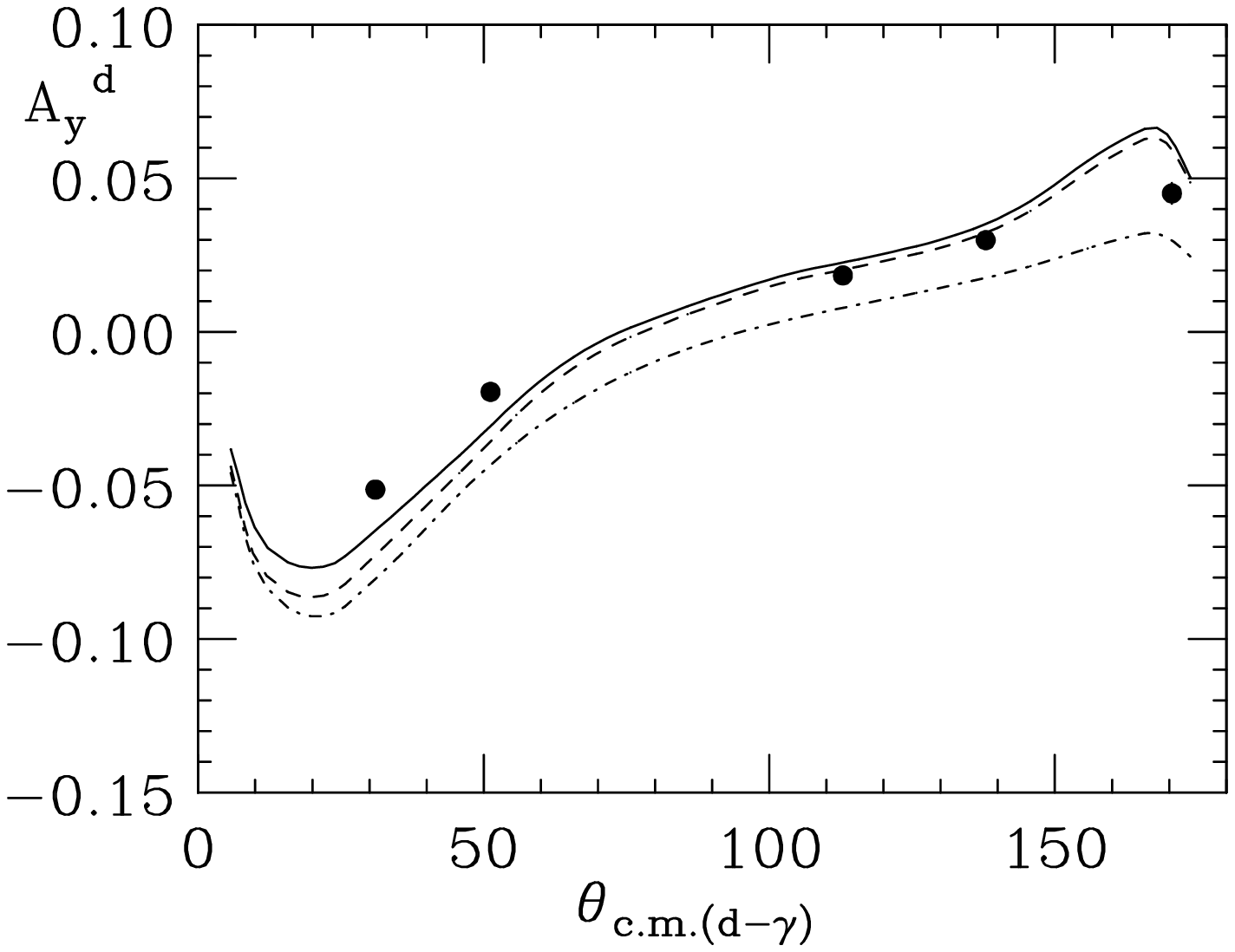}   
\includegraphics[width=0.45\textwidth]{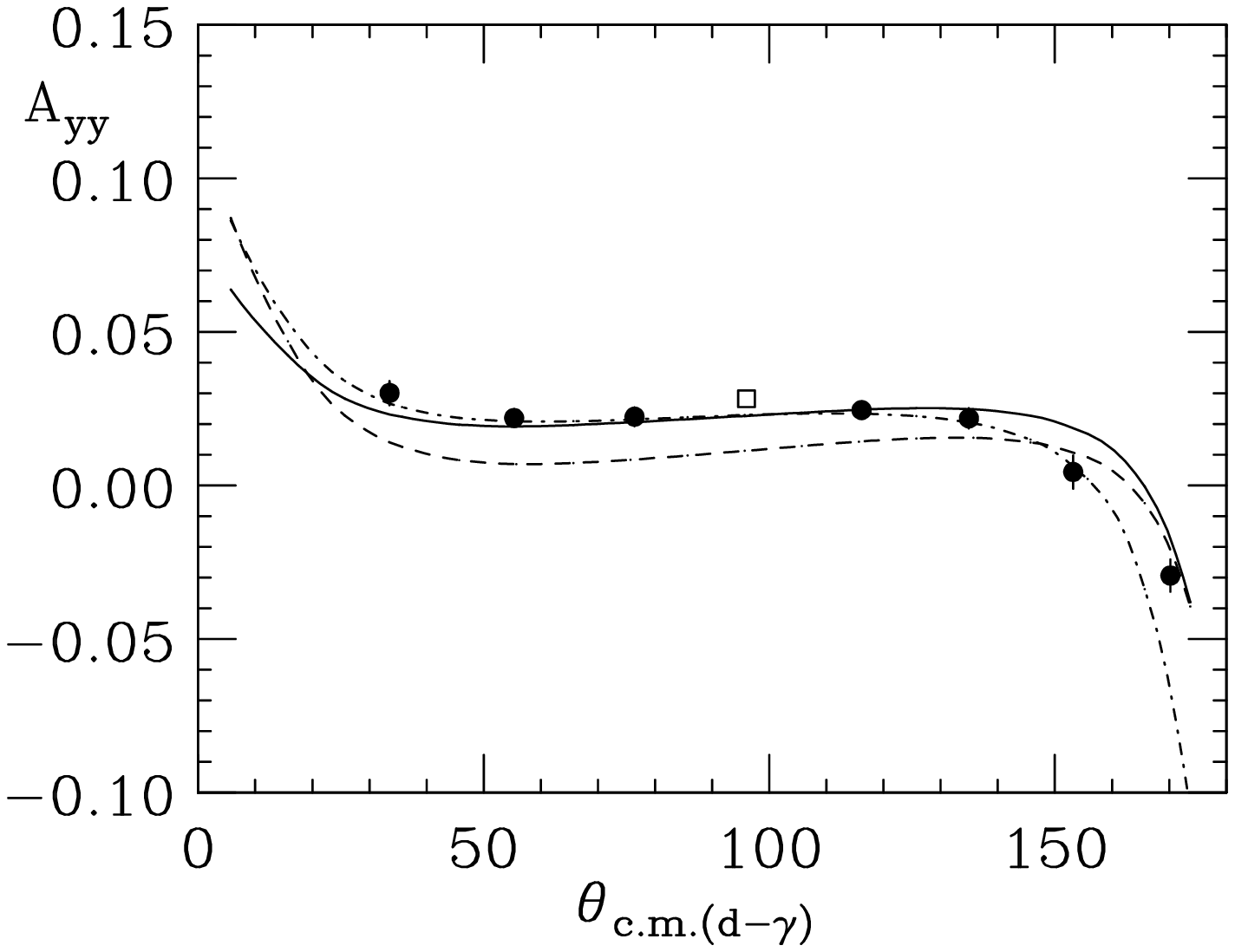}\includegraphics[width=0.45\textwidth]{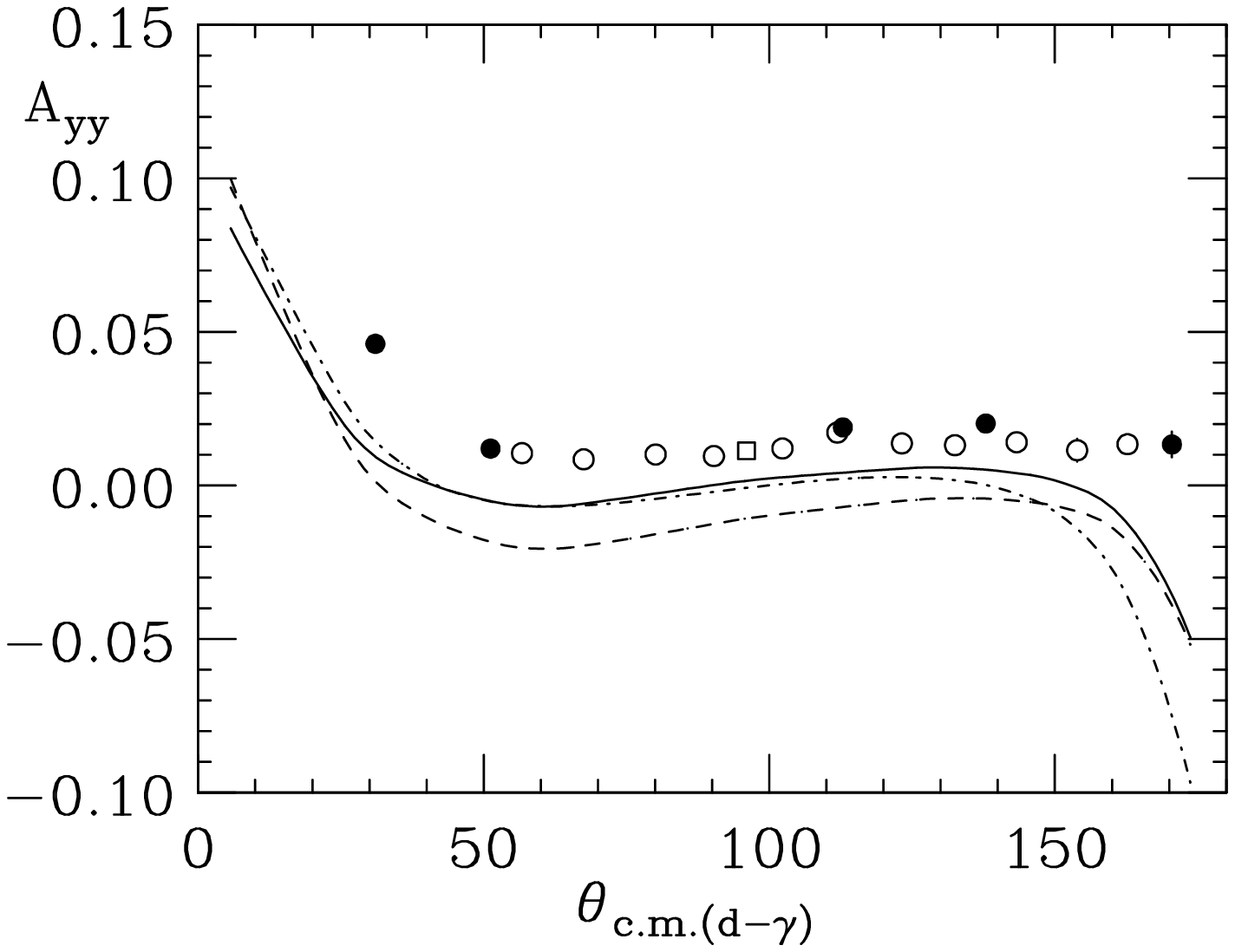}   
\caption{Same as figure {\protect{\ref{fig:1body}}}.  Here the data are compared to the calculations by Marcucci {\em et al.} with 2--body currents (dash) and with 2-- and 3--body currents (solid). The calculations by Skibinski {\em et al.} with the explicit treatment of the exchange currents are also shown (dot--dash). 
\label{fig:3body}}
\end{figure*} 

Such a more complete approach for many body currents has been employed by the PISA group.
This calculation includes $\pi$--, $\rho$--, $\omega$--, and $\sigma$--exchange currents consistent with the NN--potential as well as the currents associated with the $\rho \pi \gamma$ and $\omega \pi \gamma$ transition mechanisms and with the excitation of intermediate $\Delta$--resonances. One should note that the $\rho \pi \gamma$ and $\omega \pi \gamma$ terms are not fitted to data but used with the standard coupling constants\cite{Carlson98}. In addition, these terms give very minor effects. An alternative approach to the one by Skibinski {\em et al.} to derive the exchange currents is employed \cite{Marcucci05}. The currents are strictly consistent with the interaction potential. Thus, three--body--currents are included for calculations with the Urbana IX three--body potential. An essential aspect of the calculation is that the current conservation relation is satisfied in all calculations. 

An additional improvement of the calculations by Marcucci et al is the inclusion of the point Coulomb interaction. To account for the Coulomb interaction between the two protons is straightforward in these calculations as they are performed in configuration space. The effect is small for the analyzing powers 
of the present work but, as will be discussed below it improves the comparison between data and calculation. Figure \ref{fig:3body} shows the comparison of the data with these theoretical results.

In particular the description of the A$_y^d$ data for both energies improved with this approach in most of the angular region. If three--body--current effects are included the description also agrees with the A$_{yy}$ data at 30~MeV but still shows deviations at 45~MeV. In particular, the fall--off of the calculations in A$_{yy}$ at 45~MeV at backward angles can not be removed by many--body--currents. Deviations from the data are also present at very backward angles for the A$_y^d$ data. This suggests that additional effects are relevant for a complete description of the data. 
\begin{figure*}[htb]
\centering
\includegraphics[width=0.45\textwidth]{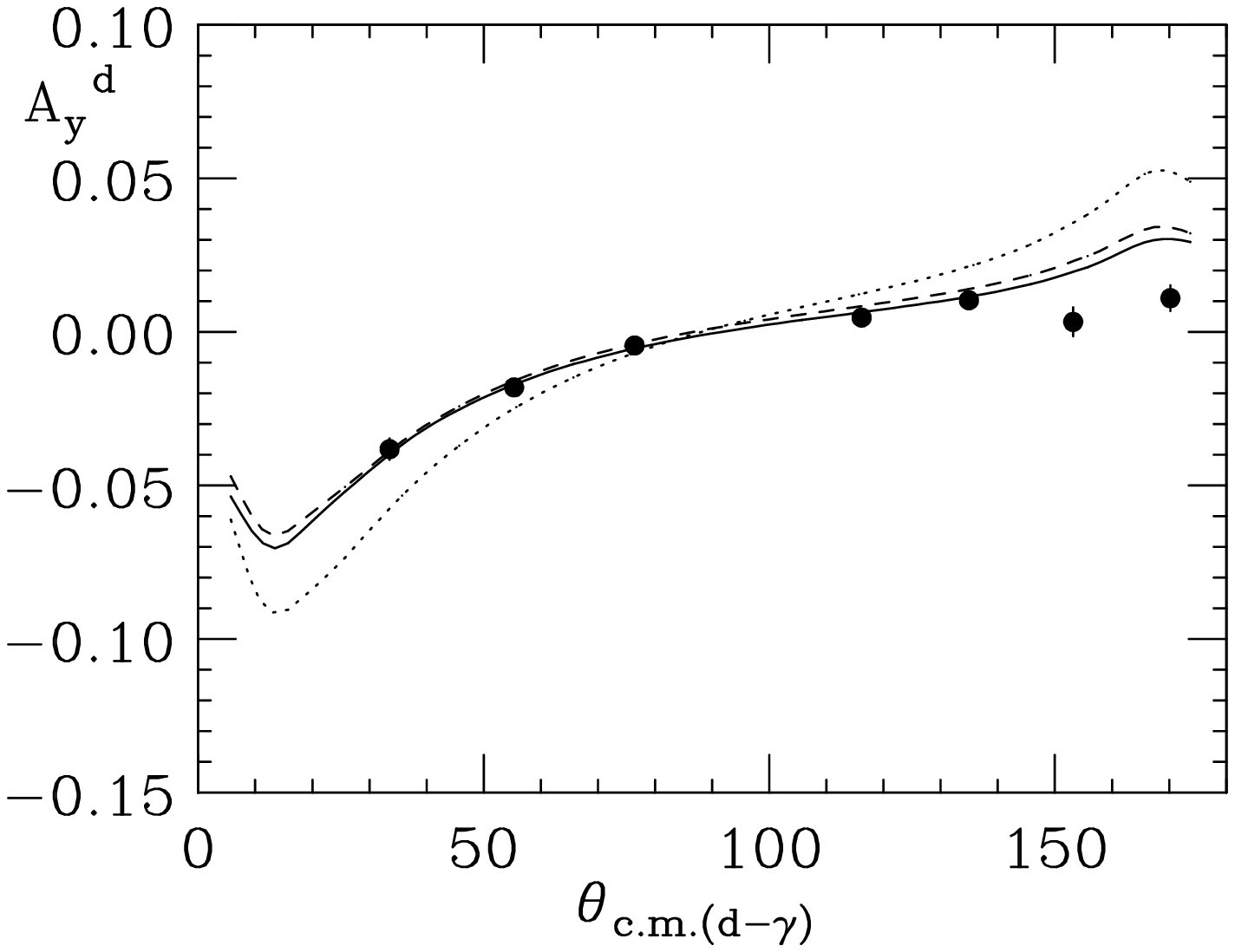}\includegraphics[width=0.45\textwidth]{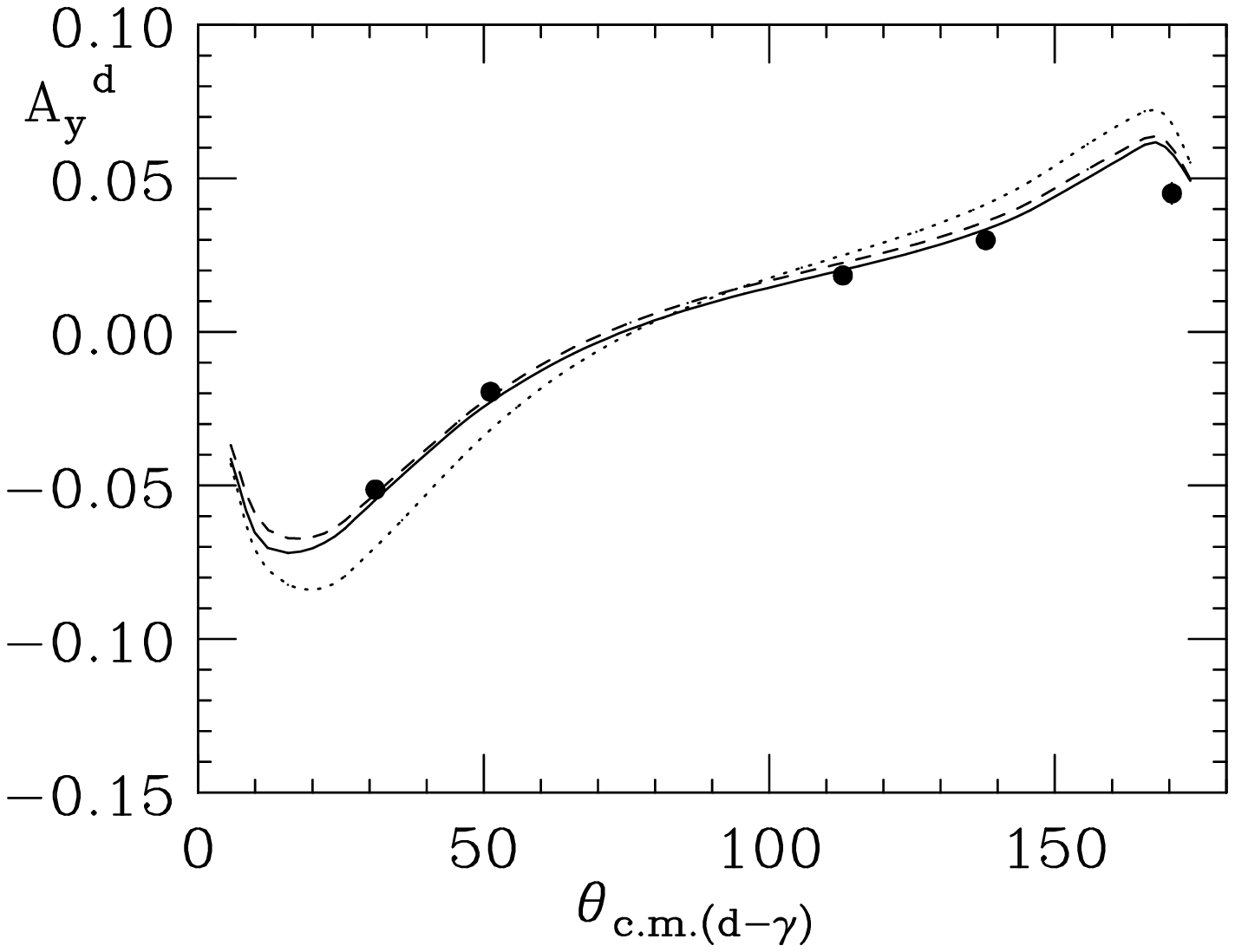}   
\includegraphics[width=0.45\textwidth]{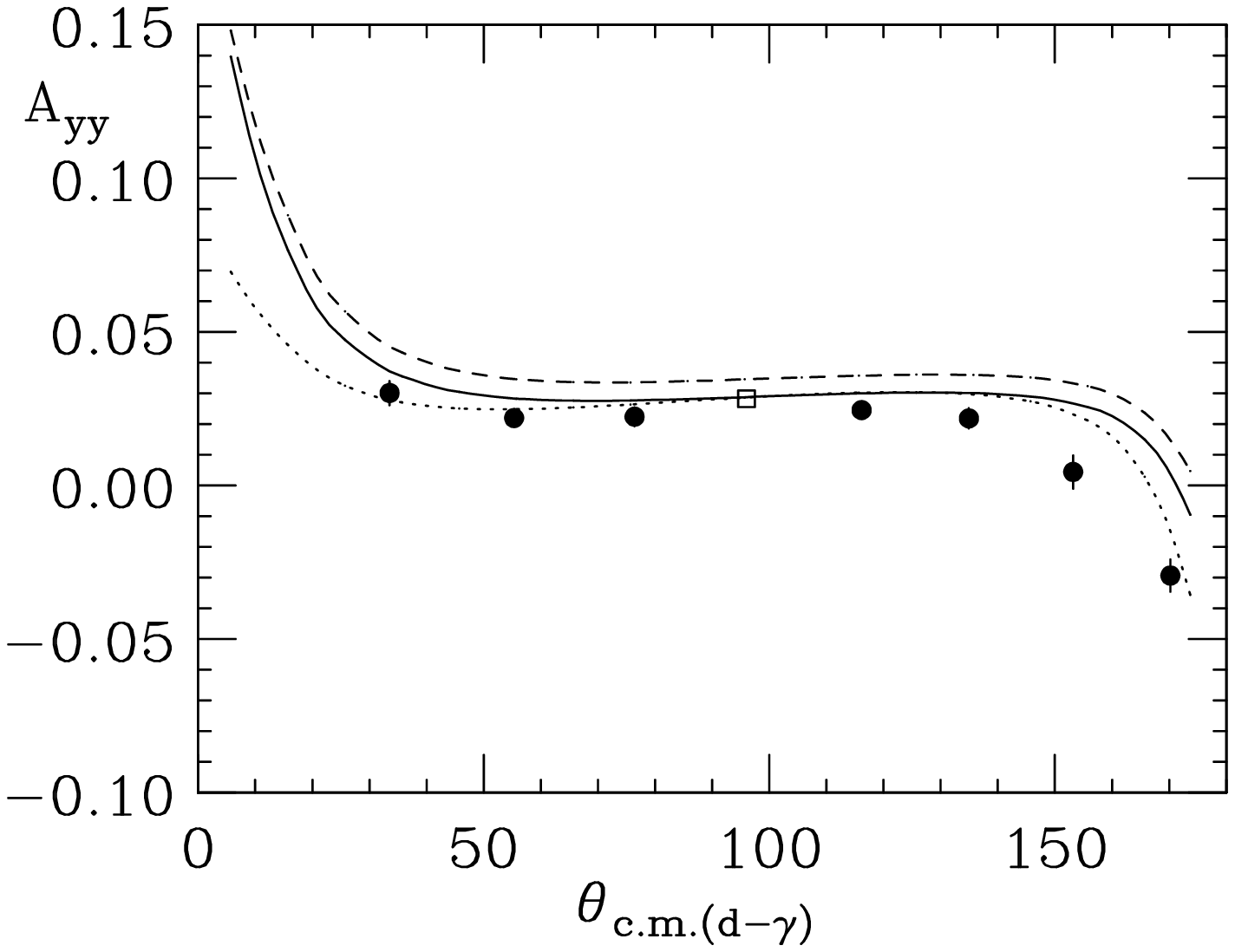}\includegraphics[width=0.45\textwidth]{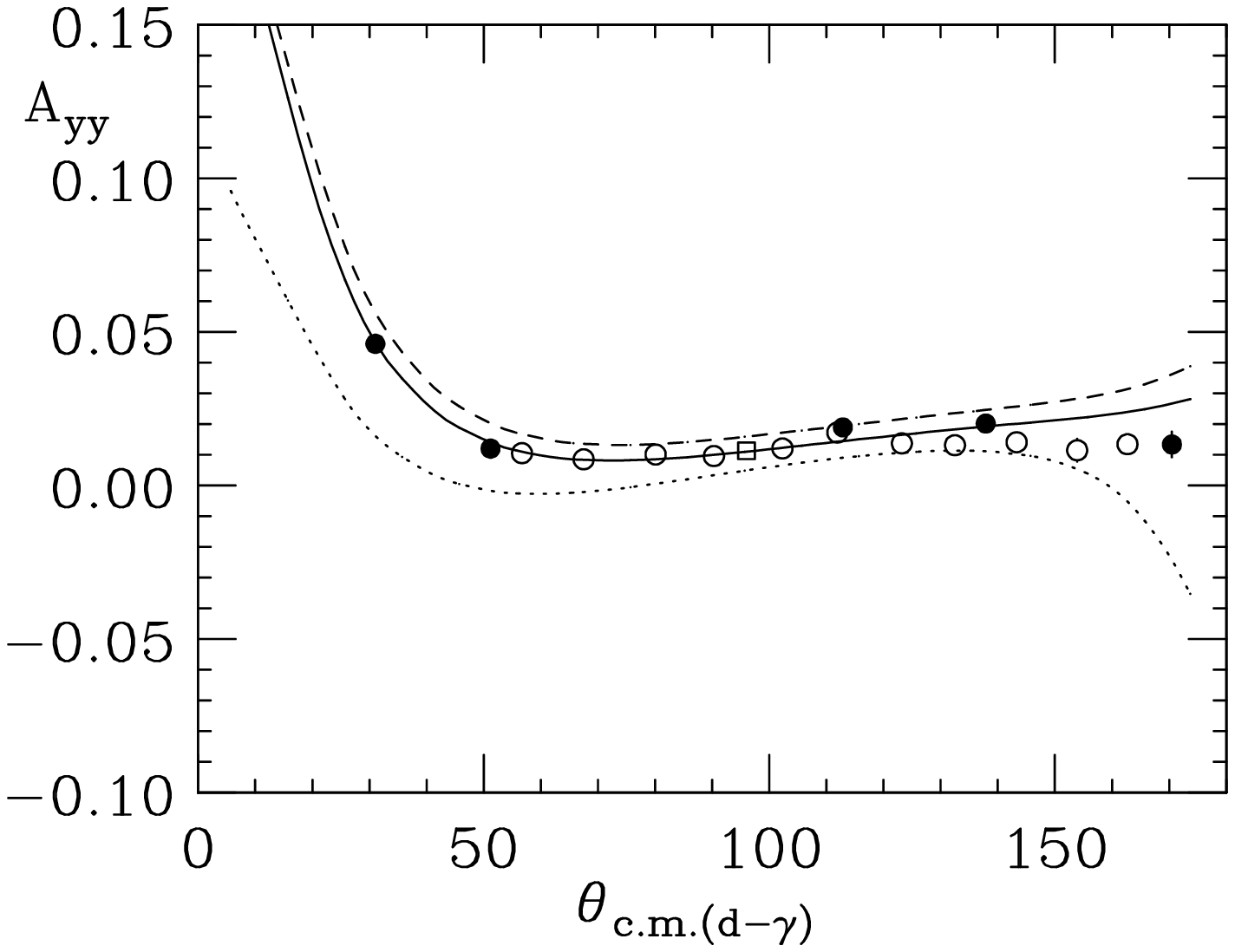}   
\caption{Same as in figure {\protect{\ref{fig:1body}}}. Comparison to the calculations by Deltuva {\em et al.} with 2--body currents (dot), plus relativistic corrections (dash), and plus added Coulomb corrections (solid).
\label{fig:rel_coul}}
\end{figure*} 

A similar discrepancy was a long standing puzzle in the 0 deg cross section of the {\em two--body} photodisintegration because relativistic effects had
been considered unimportant at these low energies. Cambi, Mosconi, and Ricci \cite{Cambi82} solved this puzzle demonstrating the importance of the relativistic spin--orbit contribution to explain the discrepancies between data and calculations.

In a $(k/m_N)$--expansion of the current operator with $k$ and $m_N$ the nucleon momentum and mass, relativistic effects of order $(k/m_N)^2$ are included in the calculation by Deltuva {\em et al.} \cite{Deltuva04b}. The two--body currents are included in this calculation using the Siegert approach without the long--wavelength approximation often applied. In contrast to the calculation by Skibinski, explicit one-- and two--body currents in the non--Siegert parts are also accounted for. 

In addition to relativistic effects, also this calculation includes the Coulomb interaction between the two protons \cite{Deltuva05}. For calculations performed in momentum space the Coulomb potential is usually omitted due to convergence difficulties. To include it a screened Coulomb potential is used, corrected for the unscreened limit using a renormalization procedure. A recent comparison with the calculations of the PISA--group discussed above demonstrates the reliability of the momentum space calculation\cite{Deltuva05a}. 
Figure \ref{fig:rel_coul} shows the data in comparison with the calculations by Deltuva {\em et al.} With this calculation the data can be reproduced over most of the angular range. The dotted line represents the results without relativistic-- and Coulomb effects. The dashed line shows the results with relativistic corrections included and the results shown with the solid line include also the Coulomb interaction. In particular the backward angle fall--off of A$_{yy}$ can be removed with the relativistic spin--orbit effect. In addition, also the backward angle deviations of the non--relativistic calculations
in A$_y^d$ are improved. 

\section{Conclusions}
Precise deuteron vector analyzing powers, A$_y^d$, and tensor analyzing powers, A$_{yy}$, have been measured for the $\mathrm{^1H(\vec{d},\gamma)^3He}$--capture reaction at two incident deuteron energies. The energies have been chosen in order to emphazise different dynamical effects. A$_{yy}$ for intermediate angles shows a maximum at 29~MeV, whereas it crosses zero around 45~MeV. 

The data have been compared to modern three--body calculations from three different groups \cite{Skibinski03,Marcucci05,Deltuva04b}. Although the groups employ very different techniques to compute the wave functions for ground-- and continuum states, the computed polarization observables agree very well with each other when only one--body--currents are included. However, large discrepancies are present between such calculations and
the data. It has been shown, that the dominant part of this discrepancy can be corrected for accounting for two-- and three--body--currents. However, it could also be shown that relativistic effects and the Coulomb interaction between the protons play a role. The agreement between calculation and data in certain angular ranges improves significantly when such effects are included.

In summary, the present precision data represent a challenging testing ground for ``state--of--the art'' three--body calculations. For a complete description of the data two-- and three--body currents as well
as relativistic-- and Coulomb effects have to be taken into account. Due to their precision the data are 
very sensitive to various small effects which usually are unobservable in cross section data. They clearly demonstrate the power of the combination of polarization data with the electromagnetic process \cite{Arenhoevel04}. 

\section{Acknowledgements}
We would like to thank A. Deltuva, P. Sauer, W. Gl\"ockle, J. Golak, R. Skibinski, H. Witala, A. Kievsky, L.E. Marcucci, and M. Vivani, for providing us with the various calculations and many valuable discussions. 
This work was supported by the Schweizerische Nationalfonds.
	
%\bibliography{/usr/users/jourdan/latex/sum2,/usr/users/jourdan/latex/sick_diff}
%\bibliographystyle{unsrt}

\end{document}